\documentclass[10pt, journal, a4paper, compsoc]{IEEEtran}

\usepackage[nocompress]{cite}
\usepackage[dvips]{graphicx}
\usepackage{fixltx2e}

\usepackage{url}

\usepackage{amssymb}
\usepackage{array}
\usepackage[table]{xcolor}
\usepackage{multirow}
\usepackage{tabularx}
\usepackage{colortbl}
\usepackage{booktabs}
\usepackage{ifthen}

\usepackage[T1]{fontenc}
\usepackage[utf8]{inputenc}

\usepackage{xspace}
\usepackage{multicol}

\usepackage{pifont}

\usepackage{ulem}

\newboolean{showcomments}
\setboolean{showcomments}{false}
\ifthenelse{\boolean{showcomments}}
{ \newcommand{\mynote}[2]{
    \fbox{\bfseries\sffamily\scriptsize#1}
    {\small$\blacktriangleright$\textsf{\textit{#2}}$\blacktriangleleft$}}}
{ \newcommand{\mynote}[2]{}}

\newboolean{showfinaltag}
\setboolean{showfinaltag}{false}
\ifthenelse{\boolean{showfinaltag}}
{\newcommand{\final}[1]{
    \fbox{\bfseries\sffamily\scriptsize FINAL}
    {\small$\blacktriangleright$\textsf{\textit{#1}}$\blacktriangleleft$}}}
{\newcommand{\final}[1]{#1}}

\newcommand{\ie}{\textit{i.e.,}}
\newcommand{\eg}{\textit{e.g.,}}

\newcommand{\etal}{\textit{et al.}}

\graphicspath{{resources/}}

\usepackage{listings}

\lstdefinelanguage{diaspec-architecture}
{morekeywords={device, attribute, extends, source, action, controller, context, from, on, void, Integer, Boolean, String, enumeration, structure, indexed, by, as},
  sensitive=true,
  morecomment=[l]{//},
  morecomment=[s]{/*}{*/},
}

\newcommand{\diagen}{Dia\-Gen\xspace}
\newcommand{\diaspec}{Dia\-Spec\xspace}
\newcommand{\diasim}{Dia\-Sim\xspace}
\newcommand{\diasuite}{Dia\-Suite\xspace{}}
\newcommand{\selfrefDiaSim}{\cite{Brun09d}}
\newcommand{\selfrefDiaSpec}{\cite{Cass09b}}

\newcommand{\myparagraph}[1]{\vspace{0.3\baselineskip}\noindent\textit{#1.}}
\newcommand{\ct}[1]{{\ttfamily #1}}

\begin{document}

\title{Towards A Tool-based\\ Development Methodology for\\ Pervasive
  Computing Applications}

\author{Damien~Cassou, Julien~Bruneau, Charles~Consel, Emilie~Balland
  \IEEEcompsocitemizethanks{\IEEEcompsocthanksitem All authors are
    affiliated with the University of Bordeaux and INRIA, FRANCE.

E-mail: first.last@inria.fr}%
\thanks{}}

\maketitle

\begin{abstract}

  Despite much progress, developing a pervasive computing application
  remains a challenge because of a lack of conceptual frameworks and
  supporting tools. This challenge involves coping with heterogeneous
  devices, overcoming the intricacies of distributed systems
  technologies, working out an architecture for the application,
  encoding it in a program, writing specific code to test the
  application, and finally deploying it.

  This paper presents a design language and a tool suite covering the
  development life-cycle of a pervasive computing application. The
  design language allows to define a taxonomy of area-specific
  building-blocks, abstracting over their heterogeneity. This language
  also includes a layer to define the architecture of an application,
  following an architectural pattern commonly used in the pervasive
  computing domain. Our underlying methodology assigns roles to the
  stakeholders, providing separation of concerns. Our tool suite
  includes a compiler that takes design artifacts written in our
  language as input and generates a programming framework that
  supports the subsequent development stages, namely implementation,
  testing, and deployment.  Our methodology has been applied on a wide
  spectrum of areas. Based on these experiments, we assess our
  approach through three criteria: expressiveness, usability, and
  productivity.

\end{abstract}


\IEEEpeerreviewmaketitle

\section{Introduction}

\IEEEPARstart{P}{ervasive} computing applications are being deployed in a
growing number of areas, including building automation, assisted
living, and supply chain management. These applications involve a wide
range of devices and software components, communicate using a variety
of protocols, and rely on intricate distributed systems technologies.
Besides requiring expertise on underlying technologies, developing a
pervasive computing application also involves domain-specific
architectural knowledge to collect information relevant for the
application, process it, and perform actions. We now review key
requirements for developing pervasive computing applications.

\myparagraph{Abstracting over heterogeneity} A pervasive computing
application interacts with \textit{entities} (\eg{} webcams and
calendars), whose heterogeneity tends to percolate in the application
code, cluttering it with low-level details. This situation requires to
raise the level of abstraction at which entities are invoked, to
factor entity variations out of the application code, and to preserve
it from distributed systems dependencies and communication protocol
intricacies.

\myparagraph{Architecturing an application} Conceptually, pervasive
computing applications collect \textit{context information} (\ie{} the
part of the environment state which is relevant for the application),
process it, and perform actions.

\final{Software development methodologies such as model-driven
engineering have been applied to design pervasive computing
applications.  A notable example is PervML~\cite{Serr10a} which relies
on the general-purpose modeling notations of UML to generate specific
programming support. However, such approaches do not provide a
conceptual framework to guide the design.  This line of work could be
taken to the next level of abstraction by offering a domain-specific
software architecturing approach.}

\myparagraph{Leveraging area-specific knowledge} Because the pervasive
computing domain includes a growing number of areas, knowledge about
each area needs to be shared and made reusable to facilitate the
development of applications. Reusability is needed at two levels.
First, it is needed at the entity level because applications in a
given area often share the same classes of entities.  Second,
reusability is needed at the application level to enable the developer
to respond to new requirements by using existing context computations
for example.

\myparagraph{Covering the application development life-cycle} Existing
general-purpose design frameworks are generic and do not fully support
the development life-cycle of pervasive computing applications. To
cover this life-cycle, a design framework specific to the pervasive
computing domain is needed. This domain-specific design framework
would improve productivity and facilitate evolution. To make this
design framework effective, the conformance between the specification
and the implementation must be guaranteed~\cite[Chap.~9]{Tayl09a}.
After the application is implemented, tools should facilitate all
aspects of its deployment.  Maintenance and evolution are important
issues for any software system~\cite{IEEE99b}.  They are even more
important in the pervasive computing domain where new entities may be
deployed or removed at any time and where users may have changing
needs. These maintenance and evolution phases should be supported by
tools.

\myparagraph{Simulation of the environment} The deployment of a
pervasive computing application requires numerous equipments to be
acquired, tested, configured, and deployed. Furthermore, some scenarios
are difficult to test because of the situations involved (\eg{} fire
in a building)~\cite{Reynolds06}. To overcome this deployment barrier,
tools should be provided to the developer to test pervasive computing
applications in a simulated environment.

\subsection*{Our contributions}

We propose an approach that covers the development life-cycle of a
pervasive computing application. It takes the form of a tool-based
methodology. The main contributions of this paper are:

\myparagraph{A design language} We introduce \diaspec{}, a design
language dedicated to describing both a taxonomy of area-specific
entities and pervasive computing application architectures.
This design language provides a conceptual framework to support the
development of a pervasive computing application, assigning roles to the
stakeholders and providing separation of concerns.  \diaspec{} raises
the level of knowledge that can be shared and reused by the
stakeholders.

\myparagraph{A tool-based methodology} We have built \diasuite{}, a
suite of tools which, combined with our design language, provides
support for each phase of the development of a pervasive computing
application, namely, design, implementation, testing, deployment,
  and evolution. \diasuite{} relies on a compiler that generates a
programming framework from descriptions written in the \diaspec{}
design language.

\myparagraph{Validation} We have successfully applied our
  methodology to a variety of applications in areas including advanced
telecommunications, home/building automation, and health-care. Based
on these experiments, we propose to assess our tool-based
methodology through three criteria: expressiveness, usability, and
productivity.

\subsubsection*{Outline.} The rest of this paper is organized as
follows. Section~\ref{sec:casestudy} presents an overview of our
tool-based methodology. Section~\ref{sec:environment-definition}
describes how a taxonomy of entities is defined using \diaspec{}.
Section~\ref{sec:architecturing} introduces the
ADL layer of \diaspec{}.
Section~\ref{sec:implementing} examines how to implement an
application, supported by a generated programming framework.
Sections~\ref{sec:testing} and~\ref{sec:deploying} discuss the
generation of support for testing and deployment, respectively.
Section~\ref{sec:chang-an-appl} presents how the design
evolution of the application can be taken into account during its
development and even after its deployment.
Section~\ref{sec:evaluating} assesses our tool-based methodology and
draws preliminary conclusions. Related work is discussed in
Section~\ref{sec:related-works} and conclusions are given in
Section~\ref{sec:conclusion}.

\section{Overview of the Methodology}
\label{sec:casestudy}

This section presents an overview of our development methodology
dedicated to pervasive computing applications. This methodology has
two main characteristics; it is (1) \textit{design-driven} and (2)
\textit{tool-based}. \final{In this section we first introduce a
  simple case study that we use throughout this paper.} Then, we show
why our methodology is design-driven through its flow of development
activities. Finally, we provide an overview of each \diasuite{} tool
that supports these development activities.

\subsection{Case Study: the Newscast Application}

We illustrate our tool-based methodology with a case study and
we introduce one of the areas involved in building
management applications, namely, the Newscast area. Newscast aims to
provide general information to users and to announce upcoming events
with respect to their preferences; an example is given by Ranganathan
\etal{}~\cite{Rang02a} for advertisement. This area requires devices to
broadcast messages (\eg{} loudspeakers and screens). As well, users
need to be identified to determine their preferences. This
identification can be achieved by various means such as short-range
badge readers. A variety of general and special-purpose information
sources can be integrated in a Newscast application. For example, a source
can consist of upcoming events. Another example of information source
can be the status of the place where the Newscast application is run,
enabling different Newscast policies (\eg{} holidays and workdays).

In our case study, our Newscast application is deployed in a school
building and has two functionalities. It first announces the upcoming
classes to the students using loudspeakers. Its second functionality
is to display customized information to the students using screens
placed at various locations in the school building. The displayed
pieces of information are the latest news about the school, as well as
the class schedules. They are displayed with respect to the interests
of the students standing near each individual screen. For example, the
information displayed on a screen depends on the spoken languages,
specialty, courses, and extracurricular activities of the students
around it.

\subsection{A Design-driven Methodology}

\begin{figure*}
  \centering
  \includegraphics[width=.9\textwidth]{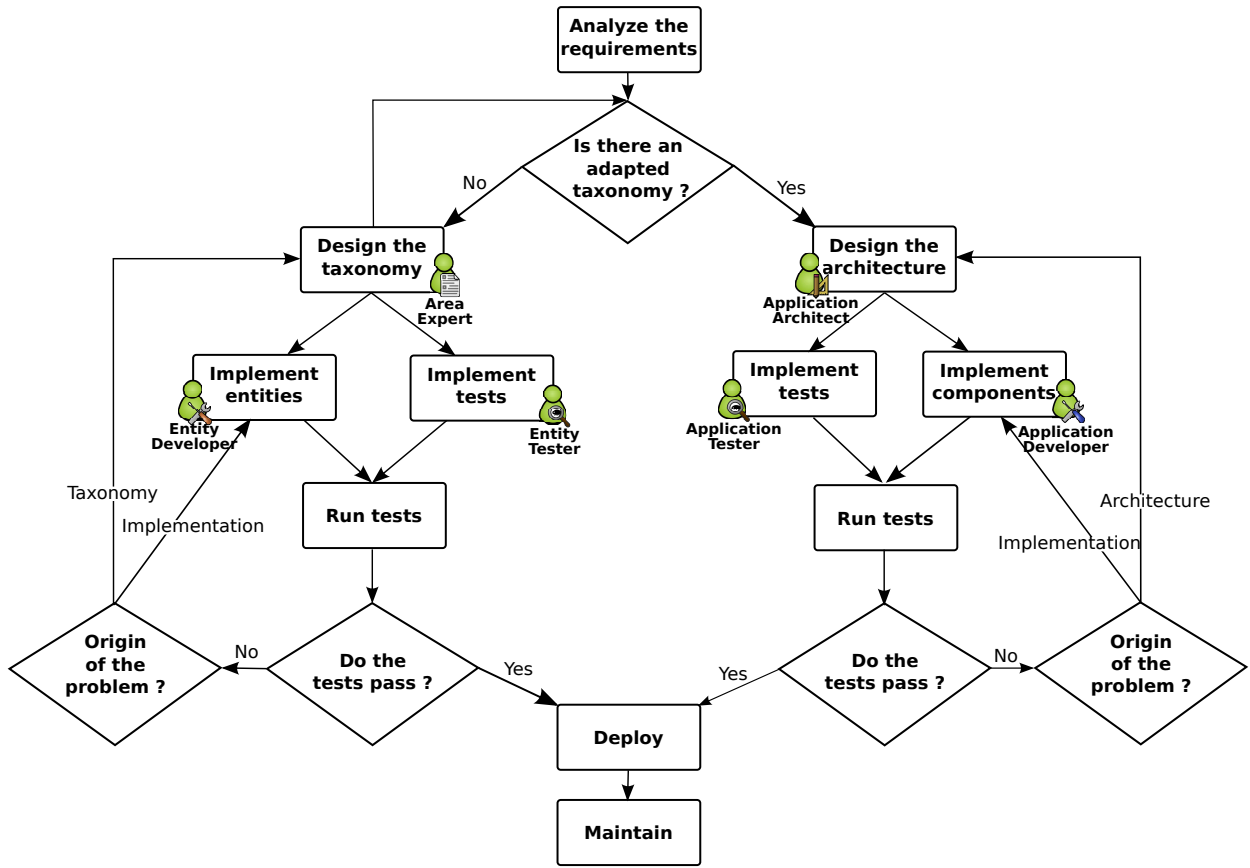}
  \caption{\label{fig:flowchart}Flowchart of the development activities
    of our tool-based methodology. Multiple inputs for an activity
    require synchronization; multiple outputs enable parallelization.}
\end{figure*}

An entity is a concept specific to the pervasive computing domain.
This concept points out the independence between (1) the development
of an entity taxonomy for an area, such as Newscast, and (2) the
development of a specific application that orchestrates elements of a
taxonomy. This independence leads to two distinct design activities:
entity taxonomy design and application design. These design
activities, as well as the subsequent implementation and testing
activities can be achieved in parallel. Figure~\ref{fig:flowchart}
outlines the development cycle associated to our methodology and
clearly illustrates the independence between these activities. In this
figure, a role is associated to each development activity. Even though
these activities are related, they can be achieved by distinct experts
in parallel, given that these experts collaborate closely.

In our development methodology, the test implementation can
start in parallel with the software component implementation as test
implementation only needs information provided by the architecture
design and comments provided by the architect. Our approach
facilitates test-driven development methodologies ({\eg} agile
software development~\cite{HighsmithFowler2001}) where the test phase
strictly precedes the implementation phase. In this way, tests guide
the developers of the application.

Along this development life-cycle, our methodology offers tools to
assist the experts for each development activity. In particular, the
specification is directly used for generating a dedicated
  programming support.

\subsection{A Tool-based Methodology}

Based on this development life-cycle and its identified roles,
Figure~\ref{fig:DevCycle} depicts how our tool suite supports each
phase of the proposed methodology:

\begin{figure}[htbp]
  \centering
  \includegraphics[width=.5\textwidth]{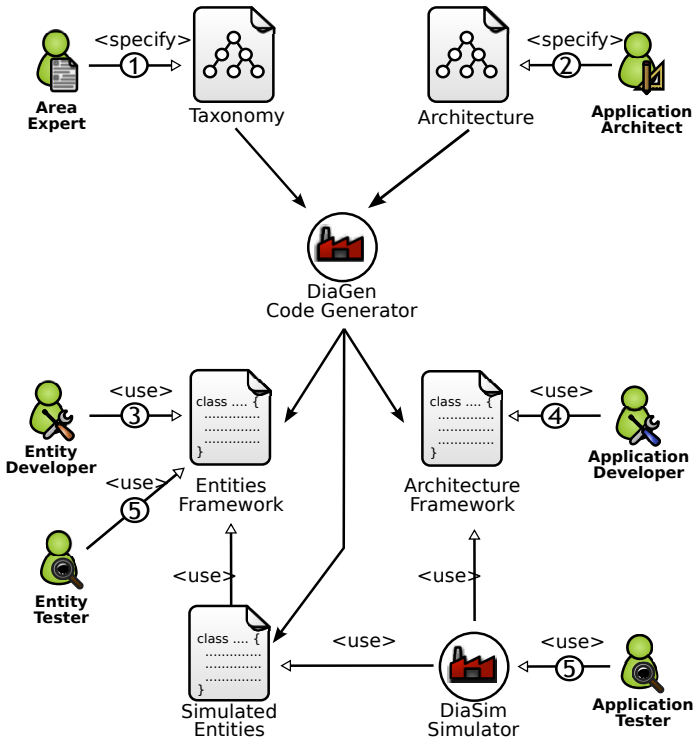}
  \caption{\label{fig:DevCycle}Development support provided by the \diasuite{} tools.}
\end{figure}

\myparagraph{Designing the taxonomy} Using the taxonomy layer of the
\diaspec{} language, an expert defines an application area through a
catalog of entities, whether hardware or software, that are specific
to a target area~(stage~\ding{192}). A taxonomy allows separation of concerns in
that the expert can focus on the high-level description of area-specific
entities.

\myparagraph{Designing the architecture} Given a taxonomy, the
architect can design and structure applications~(stage~\ding{193}). To do so, the
\diaspec{} language provides an ADL layer~\cite{Medv00a}
  dedicated to describing pervasive computing applications. An
architecture description enables the key components of an application
to be identified, allowing their implementation to evolve with the
requirements (\eg{} varying the implementation of the light management
to optimize energy consumption).

\myparagraph{Implementing entities and components} We leverage the
taxonomy definition and the architecture description to provide
dedicated support to both the entity and the application
developers~(stages~\ding{194} and~\ding{195}).
This support takes the form of a Java programming framework, generated
by the \diagen{} code generator~\selfrefDiaSpec{}. The generated programming
framework guides the developer with respect to the taxonomy
definition and the architecture description. It consists of high-level
operations to discover entities and interact with both entities and
application components. In doing so, it abstracts away from the
underlying distributed technologies, providing further separation of
concerns.

\myparagraph{Testing} \diagen{} generates a simulation support to test
pervasive computing applications before their actual deployment~(stage~\ding{196}). An
application is simulated with \diasim{}~\selfrefDiaSim{},
without requiring any code modification.  \diasim{} provides a
graphical editor to define simulation scenarios and a 2D-renderer to
monitor simulated applications. Furthermore, simulated and real
entities can be mixed. This hybrid simulation enables an application
to migrate incrementally to an actual environment.

\myparagraph{Deploying} After the testing phase, the system
administrator deploys the pervasive computing application. To this
end, a distributed systems technology is selected. We have developed a
back-end that currently targets the following technologies: Web
Services, RMI, and SIP. This targeting is transparent for the
application code. The variety of these target technologies
demonstrates that our development approach separates concerns into
well-defined layers. This separation allows to build easily new
back-ends if necessary and to smoothly apply them to already existing
applications.

\myparagraph{Maintenance and evolution} Our tool-based methodology
allows for iterative development of the taxonomy and architecture.
This approach allows changes in the taxonomy and architecture during late
phases of the cycle. 

\medskip

The next sections present in detail each one of these activities.

\section{Designing the Taxonomy}
\label{sec:environment-definition}

To cope with the growing number of application areas,
\diaspec{}\footnote{The DiaSpec grammar can be found on the website
\url{http://diasuite.inria.fr/}} offers a taxonomy language dedicated
to describing classes of entities that are relevant to the target
application area. An entity consists of sensing capabilities,
producing data, and actuating capabilities, providing actions.
Accordingly, an entity description declares a data source for each one
of its sensing capabilities. As well, an actuating capability
corresponds to a set of method declarations. An entity declaration
also includes attributes, characterizing properties of entity
instances (\eg{} location, accuracy, and status). Entity declarations are
organized hierarchically allowing entity classes to inherit
attributes, sources, and actions.

An extract of the taxonomy for the Newscast area is shown in
Figure~\ref{listing:newscast-taxo}. Entity classes are introduced by
the \textbf{device} keyword. Note that the same keyword is used to
introduce both software and hardware entities.  

To distinguish entity instances, attributes are introduced using the
\textbf{attribute} keyword. Attributes are used as area-specific
values to discover entities in a pervasive computing environment. They
also allow the tester and the system administrator to discriminate
entity instances during the simulation and deployment phases. For
example, hardware entities of our taxonomy extend the
abstract \ct{LocatedDevice} entity that introduces the \ct{area}
attribute.

The sensing capabilities of an entity class are declared by the
\textbf{source} keyword. For example, the \ct{BadgeReader} entity
defines two data sources: \ct{badge\-Detected} and
\ct{badge\-Disappeared}
(lines~\ref{newscast-taxo:badgeReader-badgeDetected-source}
and~\ref{newscast-taxo:badgeReader-badgeDisappeared-source}).
Sometimes, retrieving a data source requires a parameter. For example,
the \ct{profile} data source of \ct{ProfileDB} entity maps a badge
identifier to a user profile; in this case, the source needs to be
parametrized by a badge identifier
(line~\ref{newscast-taxo:profileDB-source}). Such parameters are
introduced by the \ct{indexed by} keyword.

Actuating capabilities are declared by the \textbf{action} keyword. As
an example, the \ct{Loud\-Speaker} declaration defines
the \ct{Play} action interface to be invoked by an application to play
a message on loudspeakers (line~\ref{newscast-taxo:loudspeaker-play}).
The play interface is defined independently in
lines~\ref{newscast-taxo:play-start} to~\ref{newscast-taxo:play-end}.
All hardware entities inherit the \ct{OnOff} action from the
\ct{SwitchableDevice} entity
(lines~\ref{newscast-taxo:switchableDevice-start}
to~\ref{newscast-taxo:switchableDevice-end}).

\begin{figure}[htbp]
  \centering
\begin{lstlisting}[language=diaspec-architecture]
device NewsProvider {
  source news as News indexed by topic as Topic;
}

device ScheduleDB {
  source todaySchedule as Schedule;
}

device BuildingStatus {
  source state as BuildingState;
}

device ProfileDB {
  source profile as UserProfile indexed#~#by#~#badge#~#as#~#String; #\label{newscast-taxo:profileDB-source}#
}

device SwitchableDevice {                       #\label{newscast-taxo:switchableDevice-start}#
  action OnOff;                                 #\label{newscast-taxo:switchableDevice-onoff-action}#
}                                               #\label{newscast-taxo:switchableDevice-end}#  

device LocatedDevice extends SwitchableDevice { #\label{newscast-taxo:locatedDevice-start}#
  attribute area as Area;                       #\label{newscast-taxo:locatedDevice-area-attribute}#
}                                               #\label{newscast-taxo:locatedDevice-end}#  

device BadgeReader extends LocatedDevice { #\label{newscast-taxo:badgeReader-start}#
  source badgeDetected as String;         #\label{newscast-taxo:badgeReader-badgeDetected-source}#
  source badgeDisappeared as String;      #\label{newscast-taxo:badgeReader-badgeDisappeared-source}#
}                                          #\label{newscast-taxo:badgeReader-end}#

device Screen extends LocatedDevice {
  attribute brightness as Integer;
  action Display;
}

device LoudSpeaker extends LocatedDevice {
  action Play; #\label{newscast-taxo:loudspeaker-play}#
}

action Play { #\label{newscast-taxo:play-start}#
  play(message as Audio);
} #\label{newscast-taxo:play-end}#

action Display {
  display(information as Information);
}

action OnOff {                                  #\label{newscast-taxo:onoff-start}#
  #on#();                                         #\label{newscast-taxo:onoff-on}#
  off();                                        #\label{newscast-taxo:onoff-off}#
}                                               #\label{newscast-taxo:onoff-end}#  

enumeration BuildingState {OPEN, CLOSE}
enumeration Topic {SPECIALTY, COURSES, EXTRACURRICULAR}
structure Area { ... }
structure UserProfile {
  name as String;
  language as Language;
  department as Department;
  ...
}
...
\end{lstlisting}
  \caption{Extract of the Newscast application
    taxonomy. DiaSpec keywords are printed in \textbf{bold}.}
  \label{listing:newscast-taxo}
\end{figure}

The taxonomy layer of \diaspec{} is domain specific in that it
offers constructs that map into concepts that are essential to the
pervasive computing domain. This is illustrated by the \textbf{source}
and \textbf{action} constructs that directly correspond to the sensing
and actuating concepts. As such, our taxonomy layer offers an
  abstraction layer between the entity implementation and the
  application logic. Indeed, on the one hand,
the entity developer takes an entity declaration as a specification
to which an entity implementation must conform. On the other hand, the
application architect can construct its specification on top of this
set of entity declarations, abstracting over the heterogeneity of
these entities.

We now present the architectural layer of the \diaspec{} language,
which is built on top of this taxonomy layer.

\section{Architecturing an Application}
\label{sec:architecturing}

\begin{figure}
  \centering
  \includegraphics[width=1\linewidth]{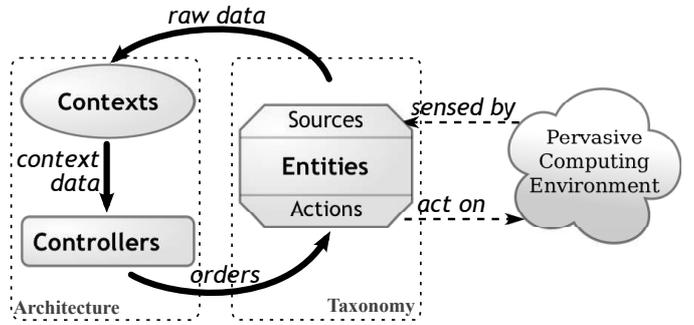}
  \caption{\label{fig:style-architectural}Architectural pattern of a
    pervasive computing application}
\end{figure}

The \diaspec{} language provides an ADL layer to define application
architectures. This layer is dedicated to an architectural pattern
commonly used in the pervasive computing domain~\cite{Dey01b,Chen02b}.
This architectural pattern is illustrated in
Figure~\ref{fig:style-architectural}. It consists of \textit{context
components} fueled by sensing entities. These components process
gathered data to make them amenable to the application needs.  Context
data are then passed to \textit{controller components} that trigger
actions on entities.

Following this architectural pattern, the ADL layer of \diaspec{}
allows the context and controller components to be defined and the
corresponding data-flow to be specified. Their definition depends on a
given taxonomy, specified in the previous step of our methodology.
Describing the application architecture allows to further specify a
pervasive computing application, making explicit its functional
decomposition.

\begin{figure}[ht]
    \centering
    \includegraphics[width=1\linewidth]{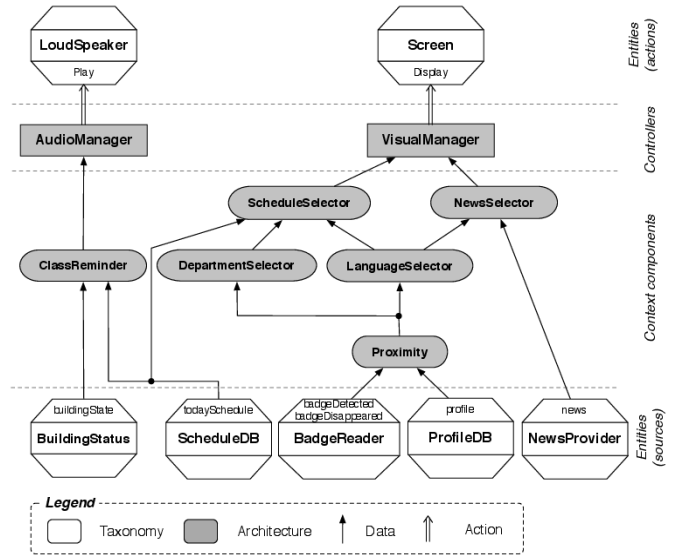}
    \caption{Layered architecture of the Newscast case study. Each
      component is described in
      Table~\ref{tab:components_newscast}.}
    \label{fig:newscast-dataflow}
\end{figure}

\newcommand{\ccolor}{\cellcolor[gray]{.8}}
\renewcommand{\tabularxcolumn}[1]{>{\arraybackslash}m{#1}}
\newcommand{\mrow}[2]{%
\multirow{#1}*{%
\begin{minipage}[t]{6em}%
\centering \arraybackslash \textbf{#2}%
\end{minipage}\arraybackslash}}

\newcommand{\mentity}[2]{\mrow{#1}{\textbf{Entity} \textit{\textnormal{(#2)}}}}

\begin{table*}
 \centering \scriptsize{}
 \begin{tabularx}{\linewidth}{c>{\tt}cX}
\toprule
\textbf{Type}        & \textrm{\textbf{Component}} & \textbf{Responsibility}                                                                                                             \\
\midrule
\mentity{5}{sensing} & BuildingStatus              & Provides the status of the building.                                                                                                \\
~                    & \ccolor{}ScheduleDB         & \ccolor{}Provides the schedule of each school department.                                                                           \\
~                    & BadgeReader                 & Reads badges close to it.                                                                                                           \\
~                    & \ccolor{}ProfileDB          & \ccolor{}Associates badge IDs to user profiles.                                                                                     \\
~                    & NewsProvider                & Provides news concerning the school.                                                                                                \\
\midrule
\mrow{7}{Context}    & ClassReminder               & Interprets the schedule of the day to only provide the classes that start in 10 minutes.                                            \\
~                    & \ccolor{}Proximity          & \ccolor{}Maps detected badge IDs into user profiles.                                                                                \\
~                    & DepartmentSelector          & Aggregates the user profiles that surround a screen and notifies when the most representative department around the screen changes. \\
~                    & \ccolor{}LanguageSelector   & \ccolor{} Selects the most representative language. (Same as the {\tt DepartmentSelector})                                          \\
~                    & ScheduleSelector            & Selects the schedule to display on a screen depending on the students surrounding the screen.                                       \\
~                    & \ccolor{}NewsSelector       & \ccolor{}Selects the news to display depending on the language of the students surrounding a screen.                                \\
\midrule
\mrow{2}{Controller} & AudioManager                & Plays messages on loudspeakers to inform about the next classes.                                                                    \\
~                    & \ccolor{}VisualManager      & \ccolor{}Displays customized news and schedules on screens.                                                                         \\
\midrule
\mentity{2}{actuating}
                     & Loudspeaker                 & Plays messages aloud.                                                                                                               \\
~                    & \ccolor{}Screen             & \ccolor{}Displays visual information.                                                                                               \\
\bottomrule
 \end{tabularx}
 \caption{Components of the Newscast application}
 \label{tab:components_newscast}
\end{table*}

We illustrate the ADL layer of \diaspec{} with the Newscast
application of our case study. The overall architecture of this
application is displayed in Figure~\ref{fig:newscast-dataflow} and all
components are described in
Table~\ref{tab:components_newscast}. At the bottom of this
  figure are the entity sources, as described in
the taxonomy. The layer above consists of the context components fueled by
entity sources. These components filter, interpret, and aggregate these
data to make them amenable to the application needs. Above
the context layer are the controller components that receive application-level data from
context components and determine the actions to be triggered on
entities. At the top of Figure~\ref{fig:newscast-dataflow}
are the entity actuators receiving actions from controller components.

\begin{figure}[htbp]
  \centering
  \begin{lstlisting}[language=diaspec-architecture]
context ClassReminder as Reminder {
  source todaySchedule from#~#ScheduleDB;
  source state from BuildingStatus;
}

context Proximity as UserProfile[] indexed#~#by#~#area#~#as Area { #\label{newscast-archi:proximity-start}#
  source badgeDetected, badgeDisappeared from#~#BadgeReader; #\label{newscast-archi:proximity-badgeReader}#
  source profile from ProfileDB;                 #\label{newscast-archi:proximity-profileDB}#
}                                          #\label{newscast-archi:proximity-end}#

context LanguageSelector as Language indexed#~#by#~#area#~#as Area {
  context Proximity;
}

context DepartmentSelector as Department indexed by area as Area {
  context Proximity;
}

controller VisualManager {                     #\label{newscast-archi:visualManager-start}#
  context NewsSelector;
  context ScheduleSelector;
  action Display on Screen;                   #\label{newscast-archi:visualManager-screen}#
  [...]
}                                                      #\label{newscast-archi:visualManager-end}#
\end{lstlisting}
  \caption{Extract of the Newscast application architecture}
  \label{listing:newscast-architecture}
\end{figure}

We now present the salient features of the \diaspec{} ADL by examining
a description fragment from the Newscast architecture and devoted to
the display of the class schedules (see
Figure~\ref{listing:newscast-architecture}). At the bottom of this
architecture is the badge reader, declared in the taxonomy, that
detects student badges in close proximity to a screen. Badge
identifiers are sent to the \ct{Proximity} component, declared using
the \textbf{context} keyword
(lines~\ref{newscast-archi:proximity-start}
to~\ref{newscast-archi:proximity-end}). This component is responsible
for maintaining the list of students, keeping an account of students
leaving or entering the screen area. To do so, it processes three
sources of information: one for entering badges, one for leaving
badges, and one for associating badge identifiers to user profiles.
These sources are declared using the \textbf{source} keyword that
takes a source name and a class of entities. The first two sources
(line~\ref{newscast-archi:proximity-badgeReader}) are bound to the
same entity class, namely \ct{BadgeReader}.

The \ct{Proximity} component signals changes in the list of students
in close proximity to a screen. To produce these changes in a
high-level form, it maps badge identifiers into user profiles by using
the \ct{ProfileDB} source
(line~\ref{newscast-archi:proximity-profileDB}). Because each list of
user profiles published by the \ct{Proximity} context is relative to a
particular screen, the architect attaches an \ct{area} to this list
through the \ct{indexed by} keyword. This value identifies the area
surrounding the screen.

The output of the \ct{Proximity} component is used by both the
\ct{Department\-Selector} and \ct{Lan\-guage\-Selector} components. These
components respectively determine the dominating department
affiliation and spoken language of the nearby students. The
\ct{Sched\-ule\-Selector} component then combines these pieces of context
information and the \ct{Sched\-ule\-DB} source to decide what schedule
should be displayed. To process this context on a per-area basis, all
the related context components are declared as indexed by \ct{Area}.
The \ct{VisualManager} controller component declares two sources:
\ct{ScheduleSelector} and \ct{NewsSelector}
(lines~\ref{newscast-archi:visualManager-start}
to~\ref{newscast-archi:visualManager-end}). It operates a screen
and thus declares the \ct{Display} action on the \ct{Screen} entity
class with the \textbf{action} keyword.

The Newscast architecture illustrates the domain-specific nature of
the \diaspec{} ADL, providing the developer with pervasive computing
concepts. These concepts are high level, making an architecture
description concise and readable. It represents a useful artifact to
share with application developers and other stakeholders. Moreover,
the \diagen{} code generator turns the role of this artifact
from contemplative to productive, guiding the implementation of the
declared components.

\section{Implementing an Application}
\label{sec:implementing}

\diagen{} automatically generates a Java programming framework from
both a taxonomy definition and an architecture description. After
outlining the implementation of DiaGen, we briefly present a generated
programming framework. This presentation is then used to explain how a
developer implements entities and the application logic on top of that
framework.

\subsection{Programming Framework Generator}

\begin{figure}
  \centering
    \includegraphics[width=1\linewidth]{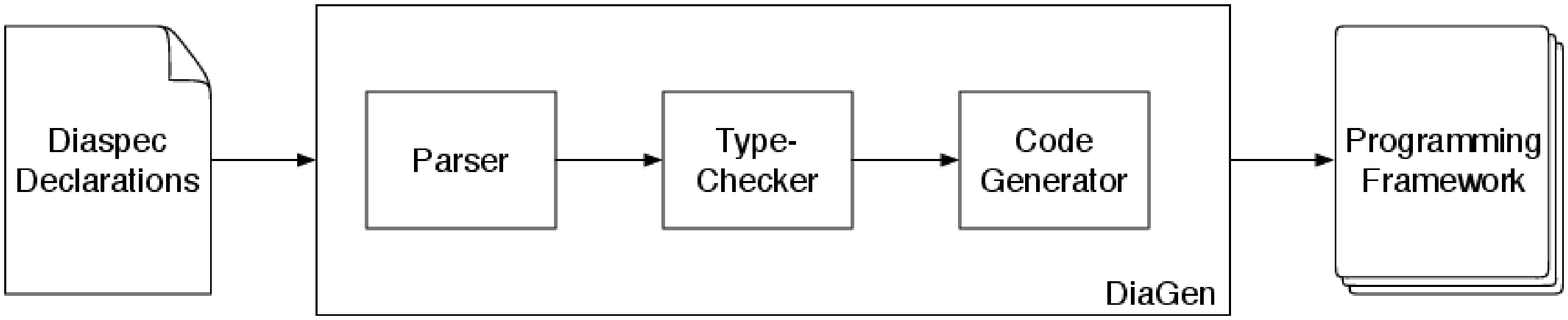}
    \caption{\label{fig:diagen}Structure of the \diagen{} compiler.} 
\end{figure}

\diagen{} generates a Java programming framework with respect to a
taxonomy definition and an architecture description. \diagen{} follows
the design of typical code generators. As illustrated in
Figure~\ref{fig:diagen}, there are three main phases: (1) the
parser, (2) the type checker, and (3) the code generator. 

The parser relies on the ANTLR\footnote{\url{http://antlr.org/}}
parser generator. Using a parser generator allows to easily
refine/extend the \diaspec{} language. The resulting Abstract Syntax
Tree (AST) is then type-checked, ensuring for example that the
inter-component communications conform to the paradigm (\eg{} a
controller cannot communicate directly with the source of an entity).
The type-checker is implemented in Java, using visitors. Finally, the
code generator is in charge of producing the programming framework
from the AST. The generator is written using the
StringTemplate\footnote{\url{http://stringtemplate.org/}} engine.
StringTemplate is a Java template engine for generating source code,
web pages, or any other formatted text output. Our suite of tools,
\diasuite{}, has been released and is available for
download.\footnote{\url{http://diasuite.inria.fr}}

\subsection{Generated Programming Framework}

A generated programming framework contains an \textit{abstract class}
for each \diaspec{} component declaration (entity, context, and
controller) that includes generated methods to support the
implementation (\eg{} entity discovery and interactions). The
generated abstract classes also include abstract method declarations
to allow the developer to program the application logic (\eg{}
triggering entity actions). 

Implementing a \diaspec{} component is done by \textit{sub-classing}
the corresponding generated abstract class. In doing so, the developer
is required to implement each abstract method. The developer writes
the application code in subclasses, not in the generated abstract
classes. As a result, in our approach, one can change the \diaspec{}
description and generate a new programming framework without
overriding the developer code. The mismatches between the existing
code and the new programming framework are revealed by the Java
compiler.

A generated programming framework also contains
\textit{proxies} to allow entities to be distributed over the network.
This is complemented by interfaces that allow the developer to
interact with entities transparently, without dealing with the
distributed systems details. Finally, a programming
  framework contains high-level support to manipulate sets of
entities easily, following the Composite design
pattern~\cite{Gamm95a}.

We now describe the process of implementing an entity, a
context component, and a controller component by leveraging a generated
programming framework. Along the way, we explain how the developer is
guided by a dedicated programming framework.

\subsection{Implementation of Entities}
\label{sec:impl-an-entity}

The compilation of an entity declaration in the taxonomy produces a
dedicated skeleton in the form of an abstract class depicted in
Figure~\ref{listing:badgeReader-generated}. We now examine the
generated support for each part of an entity declaration: attributes, sources,
and actions.

\begin{figure}[htbp]
\centering
\begin{lstlisting}[language=Java, numbers=right]
// from line  #\ref{newscast-taxo:badgeReader-start}#
public abstract class BadgeReader {

  protected BadgeReader(Area area) { #\label{badgeReader-generated:constructor}#
     super(...);
     setArea(area); #\label{badgeReader-generated:setArea-call}#
  }

  // from LocatedDevice, line #\ref{newscast-taxo:locatedDevice-area-attribute}#
  private Area area;
  public Area getArea() {return area;}
  protected void setArea(Area area) {...} #\label{badgeReader-generated:setArea}#

  // from BadgeReader, line #\ref{newscast-taxo:badgeReader-badgeDetected-source}#
  protected void setBadgeDetected(String badgeDetected) {...} #\label{badgeReader-generated:setBadgeDetected}#

  // from BadgeReader, line #\ref{newscast-taxo:badgeReader-badgeDisappeared-source}#
  protected void setBadgeDisappeared(String badgeDisappeared) {...} #\label{badgeReader-generated:setBadgeDisappeared}#

  // from SwitchableDevice, line #\ref{newscast-taxo:onoff-on}#
  public abstract void on(); #\label{badgeReader-generated:on}#

  // from SwitchableDevice, line #\ref{newscast-taxo:onoff-off}#
  public abstract void off(); #\label{badgeReader-generated:off}#
  ...
}
\end{lstlisting}
  \caption{The Java abstract class \ct{BadgeReader} generated by
    \diagen{} from the declaration of the BadgeReader entity
    (Figure~\ref{listing:newscast-taxo},
    lines~\ref{newscast-taxo:badgeReader-start}
    to~\ref{newscast-taxo:badgeReader-end})}
\label{listing:badgeReader-generated}
\end{figure}

\myparagraph{Attributes}
Entities are characterized by attributes. These attributes can be
assigned values at runtime. Attributes are managed by generated
getters and setters in the abstract class. For example, the
\ct{BadgeReader} entity has an \ct{area} attribute (inherited from
\ct{Lo\-cat\-ed\-De\-vice}, Figure~\ref{listing:newscast-taxo},
line~\ref{newscast-taxo:locatedDevice-area-attribute}) that triggers
the generation of an implemented \ct{setArea} method
(Figure~\ref{listing:badgeReader-generated},
line~\ref{badgeReader-generated:setArea}). In each subclass, the
developer will use the \ct{setArea} method to set the location of the
badge reader. The initial value for each attribute must be passed to
the generated constructor (Figure~\ref{listing:badgeReader-generated},
line~\ref{badgeReader-generated:constructor}).

\myparagraph{Sources}
An entity source produces values for context components. Support for this
propagation is generated by \diagen{}, allowing the entity developer
to invoke these methods to fuel this process. For example, from the
\ct{BadgeReader} declaration and its two sources
(Figure~\ref{listing:newscast-taxo}, lines~\ref{newscast-taxo:badgeReader-badgeDetected-source}
and~\ref{newscast-taxo:badgeReader-badgeDisappeared-source}), the
generated abstract class (Figure~\ref{listing:badgeReader-generated})
implements the \ct{set\-Badge\-De\-tected}
(line~\ref{badgeReader-generated:setBadgeDetected}) and
\ct{setBadge\-Disappeared}
(line~\ref{badgeReader-generated:setBadgeDisappeared}) methods. These
methods are to be called by a badge reader implementation.

\myparagraph{Actions}
An action corresponds to a set of operations supported by an entity.
It takes the form of a set of abstract methods implemented by
the abstract class generated for an entity declaration. Each operation
is to be implemented by the entity developer in the subclass. This
implementation bridges the gap between the declared interface and an
actual entity implementation. For example, the generated \ct{BadgeReader}
abstract class (Figure~\ref{listing:badgeReader-generated}) declares
the abstract methods \ct{on} and \ct{off}
(lines~\ref{badgeReader-generated:on}
and~\ref{badgeReader-generated:off}) that need to be implemented in
all subclasses.

The code fragment in Figure~\ref{listing:badgeReaderBluetooth} is an
implementation of a \ct{BadgeReader} entity that uses Bluetooth to
detect nearby Bluetooth devices. The \ct{BadgeReaderBluetooth}
implementation of the \ct{on} and \ct{off} methods
(Figure~\ref{listing:badgeReaderBluetooth},
lines~\ref{badgeReaderBluetooth:on-start}
to~\ref{badgeReaderBluetooth:off-end}) relies on a third-party
Bluetooth library.

\begin{figure}[htbp]
  \centering
\begin{lstlisting}[language=Java,breakatwhitespace=true]
public class BadgeReaderBluetooth extends#~#BadgeReader implements#~#BluetoothDiscoveryListener  {

  BluetoothAutoDiscovery btDiscovery;

  public BadgeReaderBluetooth(Area area) {
     super(area);
     btDiscovery = new BluetoothAutoDiscovery(this);
  }

  @Override  #\label{badgeReaderBluetooth:on-start}#
  public void on() { btDiscovery.start(); }

  @Override  #\label{badgeReaderBluetooth:off-start}#
  public void off() { btDiscovery.stop(); }  #\label{badgeReaderBluetooth:off-end}#

  // from the BluetoothDiscoveryListener interface.
  // Called by BluetoothAutoDiscovery when a new device is detected
  @Override
  public void deviceDiscovered(BluetoothDevice btDev) {
    setBadgeDetected(btDev.getAddress());
  }

  // from the BluetoothDiscoveryListener interface.
  // Called by the BluetoothAutoDiscovery when a device is not detected anymore
  @Override
  public void deviceDisappeared(BluetoothDevice btDev) {
    setBadgeDisappeared(btDev.getAddress());
  }
}
\end{lstlisting}
\caption{A developer-supplied Java implementation of a BadgeReader
  entity. This class extends the generated abstract class shown in
  Figure~\ref{listing:badgeReader-generated}. The implementation
  relies on a third party Bluetooth library: \ct{deviceDiscovered}
  and \ct{deviceDisappeared} are callback methods from the
    \ct{BluetoothDiscoveryListener} interface.}
  \label{listing:badgeReaderBluetooth}
\end{figure}

\subsection{Developing the Application Logic}

The implementation of a context or controller component also relies on
generated abstract classes. The development of the application logic
thus consists of sub-classing the generated abstract classes.

\subsubsection{Implementation of context components}

From a context declaration, \diagen{} generates programming support to develop the
context processing logic. The implementation of a context component
processes input data to produce refined data to its consumers. The
input data are either pushed by an entity source or pulled by the
context component. Both modes are provided to the developer for each
source declaration of the architecture.

The code fragment in Figure~\ref{listing:proximity} presents the
implementation of the \ct{Proximity} context (from
Figure~\ref{listing:newscast-architecture},
lines~\ref{newscast-archi:proximity-start}
to~\ref{newscast-archi:proximity-end}). This is done by extending the
corresponding generated abstract class named \ct{Proximity}.
This implementation starts by discovering all available badge readers
using the \ct{all\-Badge\-Readers} method
(line~\ref{proximity:allBadgeReaders-call}). This method is provided
by the \ct{Proximity} abstract class. The
\ct{subscribe\-Badge\-Detected} method is invoked to subscribe to the
\ct{badgeDetected} input source. Thus, the \ct{Proximity} component is
notified when a badge reader detects a new badge.

When a context component declares an input source, an abstract method
is generated in the abstract class for handling the data reception.
This method is then implemented by the developer. In
Figure~\ref{listing:proximity}, the \ct{onNew\-Badge\-Detected} method
(lines~\ref{proximity:onNewBadgeDetected-start}
to~\ref{proximity:onNewBadgeDetected-end}) illustrates such
implementation; it updates and propagates the list of profiles for a
given area when a new badge is detected. When an input source is
declared with indices, the generated abstract method takes these
indices as parameters (\textit{e.g.,} the \ct{badge} index of the
\ct{profile} input source is a parameter of the \ct{onNewProfile}
method).

The generated framework also provides a method to pull data from an
entity source. This is illustrated in Figure~\ref{listing:proximity}
where a \ct{ProfileDB} entity is first discovered
(line~\ref{proximity:allProfileDBs-call}) and then invoked to obtain
the user profile corresponding to a given badge
(line~\ref{proximity:getProfile-call}).

Finally, the generated abstract class provides methods to publish data
uniformly, whether the consumer component is a context or a controller
(line~\ref{proximity:setProximity-call}).

\begin{figure}[htbp]
  \centering
\begin{lstlisting}[language=Java, breakatwhitespace=true, numbers=left]
public class MyProximity extends Proximity {
 
  Map<Area, List<UserProfile>> profiles = new#~#HashMap<Area, List<UserProfile>>();

  @Override
  protected void postInitialize() {
    allBadgeReaders().subscribeBadgeDetected(); #\label{proximity:allBadgeReaders-call}#
    allBadgeReaders().subscribeBadgeDisappeared();
  }
  
  @Override
  public void onNewBadgeDetected(BadgeReaderProxy proxy, String badge) { #\label{proximity:onNewBadgeDetected-start}#
    Area area = proxy.getArea();
    List<UserProfile> list = getProfilesForArea(area);
    list.add(getProfile(badge));
    setProximity(list, area);                 #\label{proximity:setProximity-call}#
  }                           #\label{proximity:onNewBadgeDetected-end}#

  @Override
  public void onNewBadgeDisappeared(BadgeReaderProxy proxy, String badge) {
    // similar to onNewBadgeDetected, but removes
    // instead of adding to the list
  }

  private List<UserProfile> getProfilesForArea(Area area) {
    List<UserProfile> list = profiles.get(area);
    if (list == null) {
      list = new ArrayList<UserProfile>();
      profiles.put(area, list);
    }
    return list;
  }

  private UserProfile getProfile(String badge) {
    // gets a handler on the ProfileDB
    ProfileDBProxy profileDB = allProfileDBs().anyOne(); #\label{proximity:allProfileDBs-call}#
    
    // asks the ProfileDB about a profile for
    // the current badge
    return profileDB.getProfile(badge); #\label{proximity:getProfile-call}#
  }

  @Override
  public void onNewProfile(ProfileDBProxy proxy, UserProfile newValue, String badge) { #\label{proximity:onNewProfile-start}#
    // nothing to do here, this context is not
    // interested by notifications from the ProfileDB
  }
}
\end{lstlisting}
  \caption{A developer-supplied implementation of the Proximity context.}
  \label{listing:proximity}
\end{figure}

\subsubsection{Implementation of controller components}
\label{sec:controller}

A controller component differs from a context component in that it
takes decisions that are carried out by invoking entity actions. A
controller declaration explicitly states which entity actions it
controls. This information is used to generate an abstract class for
each controller component. This abstract class provides support for
discovering target entities and for invoking their actions. From the
declaration of the visual manager controller
(Figure~\ref{listing:newscast-architecture},
line~\ref{newscast-archi:visualManager-start}), \diagen{} generates an
abstract class named \ct{VisualManager}.
Figure~\ref{listing:visualManager} shows an implementation for this
controller. When the \ct{NewsSelector} context produces a new value,
the \ct{onNewNewsSelector} method is invoked
(line~\ref{visualManager:onNewNewsSelector}) in the
\ct{MyVisualManager} implementation. The method starts by discovering
screens present in a given area
(line~\ref{visualManager:discover}). It then displays the news on
these screens (line~\ref{visualManager:display}) by invoking the
remote method \ct{display}. This ability to discover and command
screens from the visual manager comes from the architecture
declaration (Figure~\ref{listing:newscast-architecture},
line~\ref{newscast-archi:visualManager-screen}).

\begin{figure}[htbp]
\centering
\begin{lstlisting}[language=Java, breakatwhitespace=true]
public class MyVisualManager extends VisualManager {

	@Override
	public void onNewNewsSelector(News news, Area area) {                  #\label{visualManager:onNewNewsSelector}#
		ScreenComposite screens = discover(screensWhere().area(area)); #\label{visualManager:discover}#
		screens.display(new Information(news.content));                #\label{visualManager:display}#
	}

	@Override
	public void onNewScheduleSelector(Schedule newValue, Area area) {
           // similar to onNewNewsSelector
	}

}
\end{lstlisting}
\caption{An implementation of the VisualManager controller.}
\label{listing:visualManager}
\end{figure}

After having presented the programming support given by a
generated framework, we focus on a key mechanism to cope with
dynamicity, namely, entity discovery.

\subsection{Entity Discovery}
\label{sec:entity-discovery}

\begin{figure}[htbp]
  \centering
  \includegraphics[width=1\linewidth]{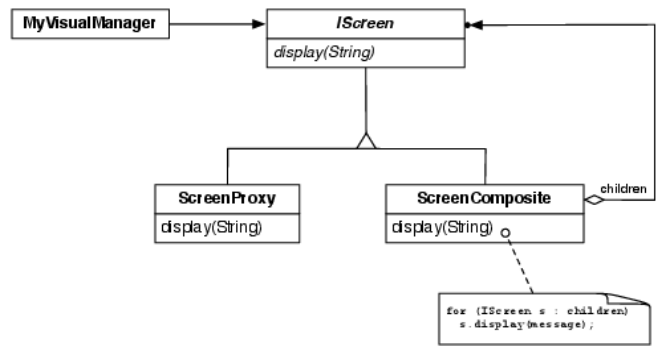}
  \caption{\label{fig:composite}Application of the composite design
    pattern to screen proxies.}
\end{figure}

Our dedicated programming framework provides support to discover
entities based on the taxonomy definition. Entity discovery returns a
collection of proxies for the selected entities. This collection is
encapsulated in a \textit{composite object} that gathers a collection
of entities~\cite{Gamm95a}. The composite design pattern applied to
screen proxies is illustrated in Figure~\ref{fig:composite}. An
example of such collection, \ct{ScreenComposite}, is returned in
line~\ref{visualManager:discover} of
Figure~\ref{listing:visualManager}. Thanks to this design pattern, the
developer can process all elements of the collection either explicitly
by using a loop, or implicitly by invoking a method of the composite,
which is part of the generated programming framework.
Line~\ref{visualManager:display} in Figure~\ref{listing:visualManager}
is an example of an implicit iteration.

To help developers express queries to discover entities, \diagen{}
generates a Java-embedded, type-safe Domain-Specific Language (DSL),
inspired by the work of Kabanov \etal{}~\cite{Kaba08a} and by fluent
interfaces introduced by Fowler~\cite{Fowl05d}. Existing works often
use strings to express queries, which defer to runtime the detection
of errors in queries. In our approach, the Java type checker ensures
that the query is well-formed at compile time. This strategy contrasts
with other works where the Java language is augmented, requiring
changes in the Java compiler and integrated development environments,
as illustrated by Silver~\cite{Vanw07a} and ArchJava~\cite{Aldr02c}.

A method suffixed by \ct{Where} is available for each device that can
be discovered. These methods return a dedicated filter object on which
it is possible to add filters over attributes associated with the
entity class. For example, the \ct{VisualManager} abstract class
defines a \ct{screensWhere} method that returns a \ct{ScreenFilter}.
This filter can be refined by adding a filter over the \ct{area}
attribute inherited by the \ct{Screen} in the taxonomy. This is done
by calling the \ct{area()} method defined in the generated
\ct{ScreenFilter} class. The parameter to this method is either an
\ct{Area} value or a logical expression. If an \ct{Area} value
  is passed, the discovered entities are those with an \ct{area}
  attribute equals to the passed value. An example of the use of a
value is given in the \ct{MyVisualManager} class shown in
Figure~\ref{listing:visualManager}. The \ct{onNewNewsSelector} method
selects screens to operate. The call to \ct{screensWhere}
(line~\ref{visualManager:discover}) restricts the selection to screens
located in the area where the news should be published.

If a logical expression is chosen, the attributes of the selected
entities hold with respect to the logical expression. A logical
expression is made of relational and logical operators. For example,
the following query selects screens that are either located in room 1
or 2:

\begin{lstlisting}[language=Java, numbers=none, basicstyle=\footnotesize\ttfamily]
Area room1, room2;
...
discover(
 screensWhere().area(or(eq(room1),eq(room2)))
);
\end{lstlisting}

\noindent
New methods can be defined to further enhance the expressiveness of
the query language. Our approach allows developers to specify filters
for more than one attribute, as shown in the following example.

\begin{lstlisting}[language=Java, numbers=none, basicstyle=\footnotesize\ttfamily{}]
Integer minSize;
Area room1;
...
discover(
 screensWhere().area(room1).size(gt(minSize))
);
\end{lstlisting}

This query selects all screens that are both in room~1
and provide a specified minimum size. Our current implementation
does not allow logical expressions across attributes. For example, it
is not possible for a query to specify that a device must have a
particular value for an attribute \textit{or} another value for another
attribute. We are working on this limitation.

This embedded DSL is both expressive and concise. It plays a key role
in enabling the developer to handle the dynamicity of a pervasive
computing environment without making the code cumbersome. 

\subsection{Interaction Modes}
\label{sec:comm-distr}

An application interacts with an entity either to
carry out an action or access data. A generated programming framework
supports the former case with the command interaction mode. The latter
case is supported by both a pull and push mode.

\myparagraph{Command} A command is a one-to-one asynchronous
interaction mode, similar to a remote procedure call. The developer
can pass arguments to a command according to signatures included in
the \diaspec{} taxonomy. 
Because a command is limited to triggering actions provided by
entities, it does not return a value, as could a remote procedure
call. An example of command invocation is given in
line~\ref{visualManager:display} of
Figure~\ref{listing:visualManager}. Instead of encoding invocation errors with a
return value, we propose a declarative approach at the architecture
level~\cite{mercadal2010}; this approach is outlined in
Section~\ref{sec:conclusion}.

\myparagraph{Pull} A context can fetch data from entities and other
contexts. To achieve these interactions, the pull mode provides a
one-to-one synchronous interaction mode with a return value. Accessing
data from an entity then consists of invoking the appropriate methods
of the entity proxy returned by the entity discovery mechanism. For
each index of the data source, a parameter is required for the method
invocation. This is exemplified by
line~\ref{proximity:getProfile-call} in
Figure~\ref{listing:proximity}.

\myparagraph{Push} This mode corresponds to the asynchronous
publish/subscribe paradigm. When a device or a context needs to push
an event (\eg{} whenever it changes), it calls a \ct{set} method
implemented in its abstract class. This is illustrated by the
\ct{MyProximity} context (Figure~\ref{listing:proximity}) that
publishes a list of user profiles located in a given area
(line~\ref{proximity:setProximity-call}) through a call to the
\ct{setProximity} method. An event value is received by all entities
that have subscribed to the event type. A subscription method is
generated in an abstract class for each entity source while
subscription to context components is automatic. The management of
subscribers and the propagation of events are supported by the
generated programming framework, further easing the development process.

\section{Testing a Pervasive Computing System}
\label{sec:testing}

As in any software engineering domain, testing pervasive computing
applications is crucial. However, this domain has specific requirements
that prevent generic testing tools from applying to pervasive
computing applications~\cite{Reynolds06}. Indeed, pervasive computing
applications interact with users and with the physical environment. Generic
software testing tools do not cope with neither the simulation of the
physical environment, nor the simulation of users in this physical
environment.

Coping with these requirements makes the testing of pervasive
computing applications challenging. In fact, only a few existing
development approaches in the pervasive computing domain address
testing: existing development approaches often assume that the system
is partially or fully deployed. However, deploying a pervasive
computing application for testing purposes can be expensive and
time-consuming because it requires to acquire, test, and configure all
equipments and software components. Furthermore, some scenarios are
difficult to test because they involve exceptional situations such as
fire.

\begin{figure*}[htbp]
  \centering
    \includegraphics[width=1\linewidth]{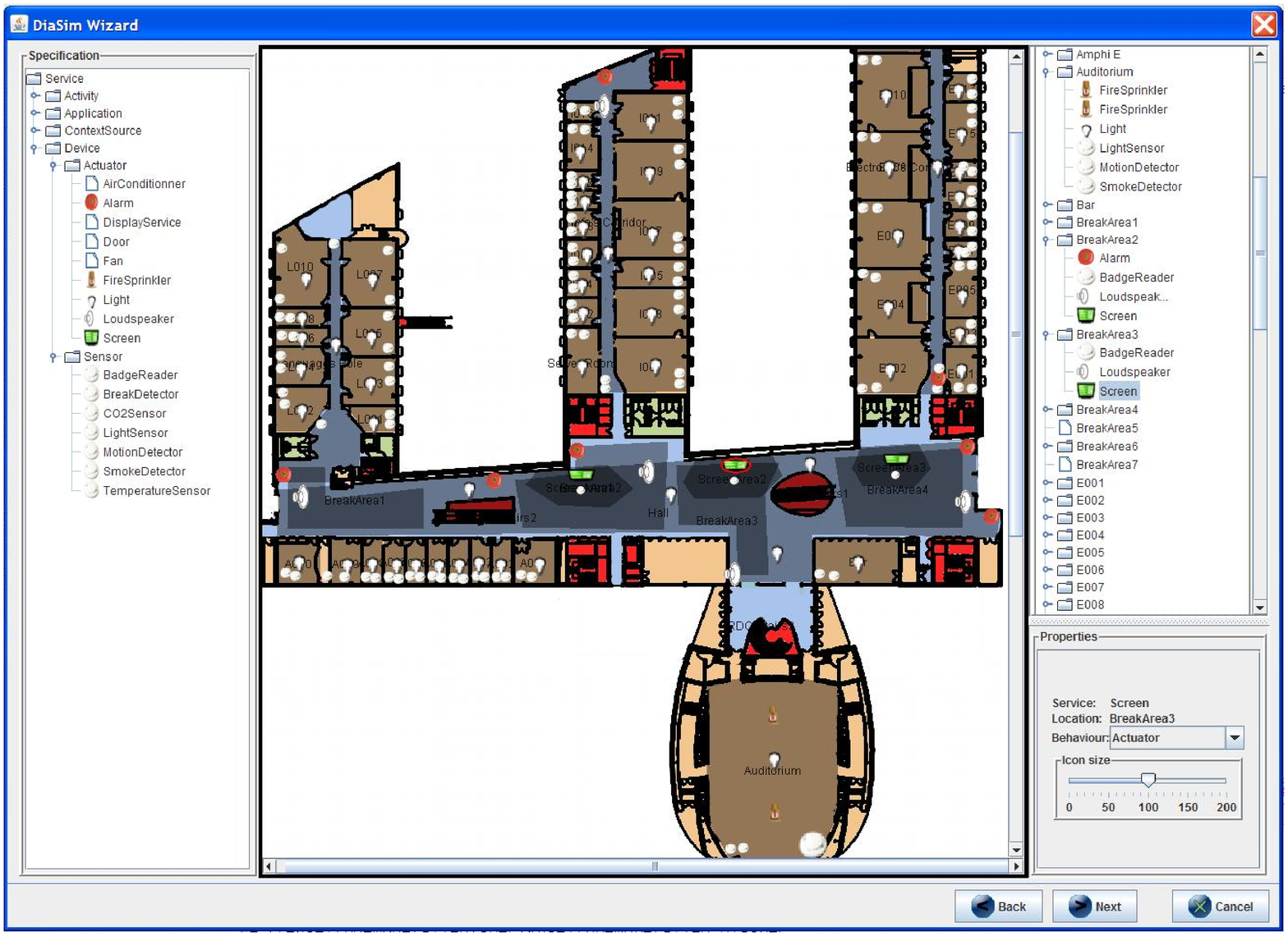}
    \caption{\label{fig:editor-enseirb}\diasim{} environment editor.
      The \diasim{} editor is parameterized by an entity taxonomy. The
      entities defined in the taxonomy are displayed on the left panel of
      the graphical user interface. The entities can be dragged and
      dropped on the central panel to add simulated entity instances
      into the simulated environment.} 
\end{figure*}

To cope with these issues, \diasuite{} includes a simulator for
pervasive computing applications, named \diasim{}. This tool is
integrated in our methodology, leveraging declarations provided at
earlier development stages. It provides support to simulate the
physical environment and execute pervasive computing applications
developed in \diasuite{}. \diasim{} leverages the abstraction layer of
the generated programming framework that allows entities to be
operated regardless of their nature (\eg{} actual or simulated). This
abstraction layer allows the simulation of these applications without
requiring any changes in the application code.

The modeling of the physical environment is described in
Section~\ref{sec:modeling_environment}. \diasim{} provides an editor to
create simulation scenarios to test pervasive computing
applications. The creation of a simulation scenario is examined in
Section~\ref{sec:defining_scenarios}. Finally, in
Section~\ref{sec:running_scenarios}, the \diasim{} runtime platform is
presented.

\subsection{Modeling the Environment}
\label{sec:modeling_environment}

The first step to simulate a pervasive computing application is to
model the physical environment. This model can be used to test
multiple pervasive computing applications. The modeling of the
physical environment is realized in a graphical editor. This editor
allows to define the layout of a physical environment, including
structural characteristics (\eg{} walls).
Figure~\ref{fig:editor-enseirb} shows the simulated school building
that we modeled for testing the Newscast application. In this example,
an area has been defined for each room, corridor, and hall of the
simulated school building. Then, using a \diaspec{} taxonomy, the
tester defines and places the simulated entity instances in the model
of the physical environment. Figure~\ref{fig:editor-enseirb}
illustrates the configuration of simulated entity instances for
testing the Newscast application. In this example, simulated
loudspeakers, screens, and badge readers are placed in the main school
hall. Simulated loudspeakers are also placed in each corridor.
Finally, simulated people are added to the simulation using the
editor. We can assign properties to each simulated person. For testing
the Newscast application, we assign a badge ID to each simulated
person. The badge ID property is then used during the simulation by
the simulated badge readers when publishing a badge detection or badge
disappearance event.

\subsection{Defining the Simulation Scenarios}
\label{sec:defining_scenarios}

As the scope of pervasive computing applications increases, so does
the range of scenarios to test. \diasim{} provides support to define
these scenarios. A simulation scenario consists of a series of
evolutions of a physical environment and simulated entity instances.
An evolution corresponds to a change in the simulated physical
environment at a specific time. During the simulation, these changes
are emitted by the simulator and processed by the simulated entity
instances. In a pervasive computing application, data sources come
from sensing stimuli from the physical environment; the collected data
are used for context processing. Simulating the environment stimuli
allows to test an application in a simulated environment. Defining the
evolution of the physical environment consists of defining these
simulated environment stimuli. The tester can either define these
environment stimuli during the edition using a stimulus library, or he
can program them using a simulation programming support. These two
sorts of support also apply for defining the behavior of simulated
entities.

\myparagraph{Stimulus and entity behavior library} During the edition,
for each stimulus needed in a simulation scenario, the tester defines
its refreshment rate and how its values evolve. As an example, the
luminosity value of an outside area could be refreshed every second
and vary every hour. To ease stimulus configuration, a library of
commonly used stimuli is provided. For instance, this library allows
to easily create sinusoidal stimuli. A simple simulation of outside
temperature and luminosity can be modeled as sinusoids. Another
library is provided to the developer with basic behaviors for
entities. For example, one behavior consists in making a sensor
periodically publish values. In the Newscast application, we used this
library of behaviors for the simulated badge readers. The chosen
behavior simply forwards simulated stimuli. Thus, when a simulated
badge reader receives a badge detection or badge disappearance
stimulus, it forwards this information to the Newscast application.

\myparagraph{Simulation programming support} Yet, new stimuli and
behaviors can be introduced; this development is facilitated by a
simulation programming support. This programming support provides a
generic \ct{StimulusProducer} class that the tester can use to create
his own stimulus producers. As well, the tester can develop his own
entity behavior by extending the abstract class provided in the
generated programming framework for this entity. In the Newscast case
study, we use this simulation programming framework to produce badge
detection and disappearance stimuli when simulated people move around
simulated badge readers. Figure~\ref{listing:agentmodel} presents the
implementation of the class that publishes badge detection and
disappearance stimuli. This class extends \ct{DiaSimAgentModel}. The
\ct{DiaSimAgentModel} class is provided by the simulation programming
framework and provides programming support for handling the simulated
people of the simulation. In this example, it is used to be notified
when a simulated agent enters or leaves the detection area surrounding
a badge reader. Two stimulus producers are created in this class:
\ct{badgeDetectedProducer} (Figure~\ref{listing:agentmodel},
line~\ref{agentmodel:badgedetected}) and \ct{badgeDisappearedProducer}
(line~\ref{agentmodel:badgedisappeared}). The simulation programming
support allows to be notified when an agent moves by implementing the
\ct{AgentListener} interface. When an agent moves, the \ct{agentMoved}
method is called (Figure~\ref{listing:agentmodel},
line~\ref{agentmodel:agentmoved}). When an agent enters the detection
area of a badge reader, a badge detection stimulus is published
(Figure~\ref{listing:agentmodel},
line~\ref{agentmodel:detectionpublish}). Likewise, when an agent
leaves the detection area of a badge reader, a badge disappearance
stimulus is published (Figure~\ref{listing:agentmodel},
line~\ref{agentmodel:disappearancepublish}).) In this example, the
detection range of the simulated badge readers is set to 5 meters
(Figure~\ref{listing:agentmodel}, line~\ref{agentmodel:range}).

\begin{figure}[htbp]
\centering
\begin{lstlisting}[language=Java, breakatwhitespace=true, numbers=right]
public class MyAgentModel extends DiaSimAgentModel implements AgentListener {

        private static int DETECTION_RANGE = 5; #\label{agentmodel:range}#
        private Map<Agent, DiaSimDevice> detectedAgents;
        private StimulusProducer badgeDetectedProducer;  
        private StimulusProducer badgeDisappearedProducer;  

        public MyAgentModel(World world) {
            super(world);
            Source badgeDetectionSource = new Source("BadgeReader", "badgeDetected", "String");
            badgeDetectedProducer = new StimulusProducer(badgeDetectionSource); #\label{agentmodel:badgedetected}#
            Source badgeDisappearanceSource = new Source("BadgeReader", "badgeDisappeared", "String");
            badgeDisappearedProducer = new StimulusProducer(badgeDisappearanceSource); #\label{agentmodel:badgedisappeared}#
            detectedAgents = new HashMap<Agent,DiaSimDevice>();
        }

	@Override
	public List<DiaSimAgent> createAgents() {       
            List<DiaSimAgent> agents = super.createAgents();
            for (DiaSimAgent a : agents) {
                agent.addAgentListener(this);
            }
            return agents;	
	}


        @Override
        public void agentMoved(Agent agent, String location) { #\label{agentmodel:agentmoved}#
            String id = agent.getProperty("badgeId");
            if (!detectedAgents.contains(agent)) {
                /* This agent has not been detected yet by a badge reader. We check if he has entered the detection area of a badge reader */
                for (DiaSimDevice d : getDevices()) {
                    if (d.getType().equals("BadgeReader") 
                           && agent.distanceFrom(d.getPosition()) < DETECTION_RANGE) {
                        detectedAgents.put(agent,d);
                        badgeDetectedProducer.publish(id,location); #\label{agentmodel:detectionpublish}#
                    }
                }
            } else {
                /* This agent has already been detected by a badge reader. We check if he has left the detection area of this badge reader */
                DiaSimDevice badgeReader = detectedAgents.get(agent);
                if (agent.distanceFrom(badgeReader.getPosition()) > DETECTION_RANGE) {
                    detectedAgents.remove(agent);
                    badgeDisappearedProducer.publish(id,location); #\label{agentmodel:disappearancepublish}#
                }
            }
       }
}
\end{lstlisting}
\caption{Implementation of the \ct{MyAgentModel} class used in the Newscast
  application simulation. This class is responsible for publishing
  badge detection stimuli when simulated people come near a badge
  reader. It is also responsible for publishing badge disappearance
  stimuli when simulated people leave the surroundings of badge readers.}
\label{listing:agentmodel}
\end{figure}

\myparagraph{Further simulation support} An ongoing work on
\diasim{} is to create physically accurate simulation of the physical
environment~\cite{bruneau2010}. This work consists of coupling
\diasim{} with the Acumen DSL~\cite{Zhu09a}. Acumen allows to describe
continuous systems with mathematical equations. For example,
temperature is a physical characteristic that can be described as a
continuous system with Acumen. The simulated environment stimuli are
computed by Acumen, allowing a physically accurate simulation of the
environment.

Once a scenario is defined, it is executed by the \diasim{} runtime
platform.

\subsection{Running the Simulation Scenarios}
\label{sec:running_scenarios}

\begin{figure*}[htbp]
\centering
\includegraphics[width=1\linewidth]{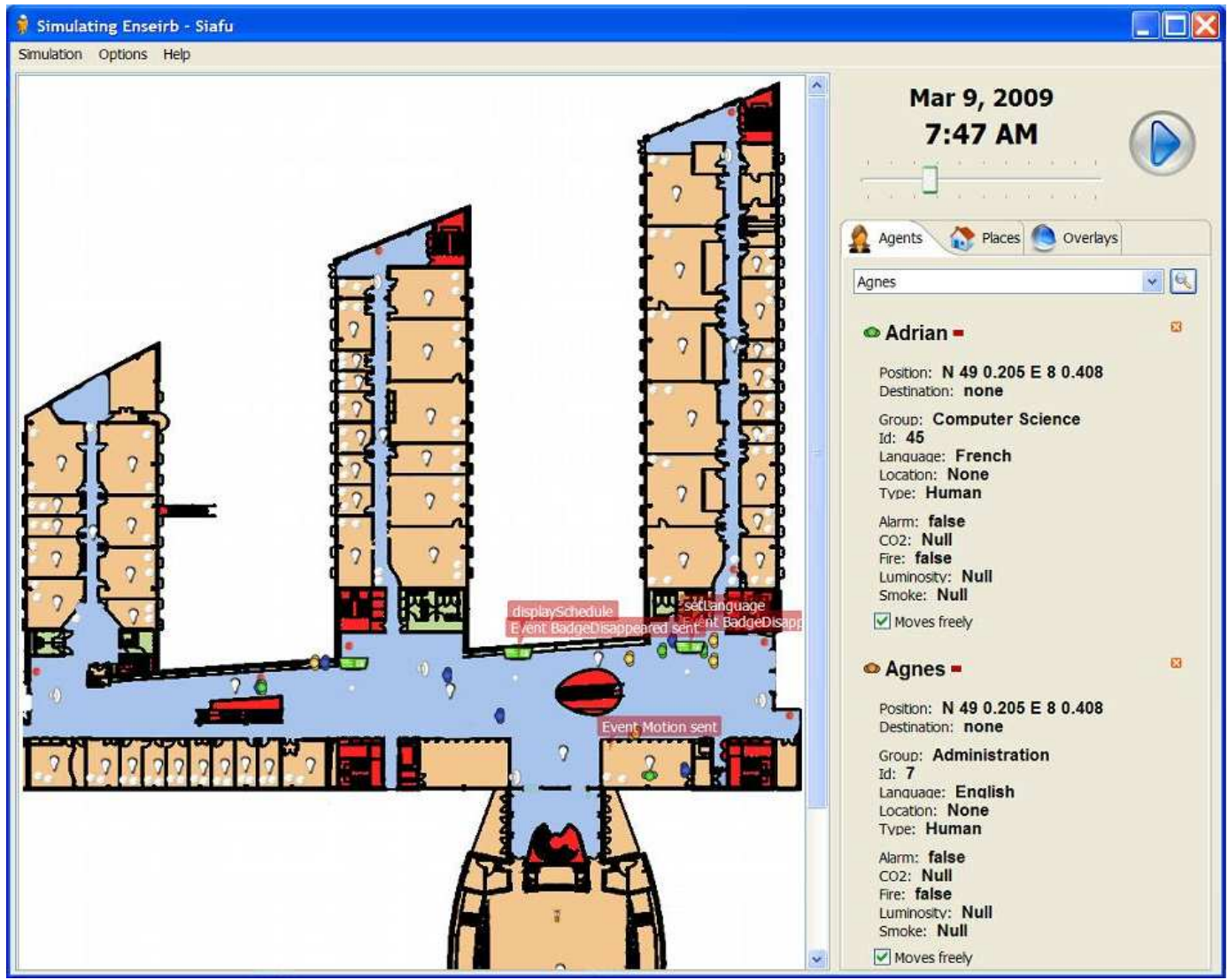}
\caption{\diasim{} scenario renderer. The simulated
  environment is displayed in the left part of the graphical user
  interface. The red popups transparently displayed above the simulated entities
  indicate that the entity has realized an interaction. More
  information about the simulated people and simulated entities can be
  found on the right of the graphical user
  interface.\label{fig:simulator-enseirb}}
\end{figure*}

Simulation scenarios are executed in the \diasim{} runtime platform. This
platform includes a 2D-graphical renderer, based on Siafu~\cite{Mart06a},
to monitor the simulation. The simulation renderer is illustrated in
Figure~\ref{fig:simulator-enseirb}. The simulated entities are
displayed in the environment representation and messages appear above
the entities when sensing or actuating is performed.

Fine-grained simulation can be achieved by manually injecting stimuli
during the simulation and plotting trajectories to move simulated
persons.

Finally, a tested application can be executed in hybrid environments,
combining simulated and real entities. Hybrid simulation is a key
feature to successfully transition to a real environment: it allows
real entities to be added incrementally in the simulation, as the
implementation and deployment progress.

\section{Deploying a Pervasive Computing System}
\label{sec:deploying}

A pervasive computing application is distributed by nature and thus
depends on the chosen distributed systems technology. When no abstraction
layer is provided to the developer, the application code embeds
distributed systems operations, creating dependencies to the
underlying technology and obfuscating the code.

Our tool-based methodology makes it possible for application code to
abstract away from the underlying distributed systems technologies,
deferring to the \diasuite{} back-end the mapping to a particular
platform. This strategy makes the application code portable across
distributed systems technologies without any change in the
implementation. When the pervasive computing application is deployed,
a distributed systems technology is selected. Four distributed systems
technologies are currently offered, targeting Web
Services~\cite{WS04a}, SIP\footnote{SIP stands for {Session Initiation
    Protocol}~\cite{Rose02a}. It is a de facto standard for modern
  telephony.}, and RMI~\cite{Down98a}. Each of these technologies
provides specific features and mechanisms with various benefits for
the development of pervasive computing applications. For example, RMI
is well-suited for testing because it only requires a lightweight
infrastructure. SIP is well-suited for pervasive computing systems
that revolve around telephony. We are working on a new back-end
mechanism that allows the entities to declare which distributed
systems technology they support. This back-end will make it possible
to mix several communication protocols in the same application,
depending on what is supported by the entities. Finally, deployment
currently requires writing Java code to instantiate the needed
entities. We believe this could be made easier by leveraging an
existing deployment technology such as OSGi~\cite{OSGi}. In
particular, we plan to generate OSGi bundles for each entity to let
OSGi handle entity life-cycles.

\section{Maintenance and Evolution}
\label{sec:chang-an-appl}

\begin{figure*}
  \centering
  \includegraphics[width=\linewidth]{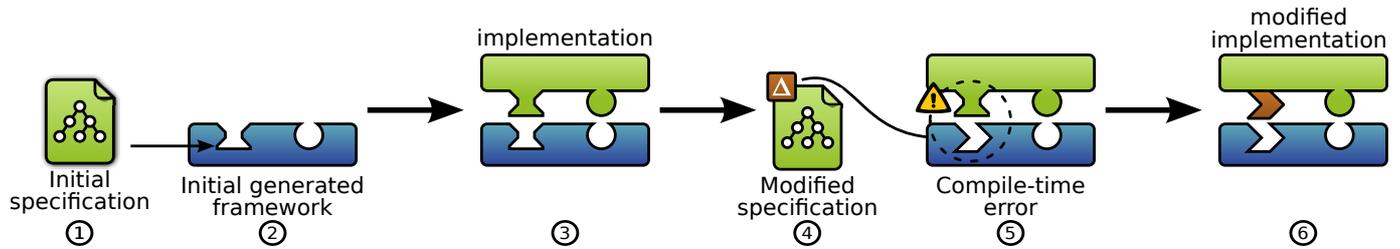}
  \caption{\label{fig:iterativeProcess} Changes in the specifications
    are possible even after framework generation. They often trigger Java
    compile-time errors that IDEs will report. In stage \ding{192}, the
    architect writes an initial specification. In stage \ding{193}, a
    programming framework is generated from the specification and a
    developer starts implementing on top of it in stage
    \ding{194}. In stage \ding{195}, the architect modifies the
    specification. A new programming framework is generated in
    stage \ding{196}. This new framework replaces the previous one; this
    triggers Java compile-time errors in existing implementations. In
    stage \ding{197}, the developer is guided by the compile-time errors to
    apply the changes of the architect to the implementation.}
\end{figure*}

Maintenance and evolution are important parts of the development of
any software system~\cite{IEEE99b}. It is even more important in the
pervasive computing domain where new entities may be deployed or
removed at any time and where users may have changing needs. To cope
with maintenance and evolution of such applications, our tool-based
methodology allows an iterative development of a pervasive
computing application. This is illustrated in
Figure~\ref{fig:iterativeProcess}.

\subsection{Evolution of the Taxonomy}

The taxonomy is likely to change in a stage subsequent to the
  programming framework generation (\eg{} during the application implementation or after
deployment). Applying these changes requires the programming framework to be regenerated. 
When the newly-generated programming framework is supplied
to the developers, the IDE automatically points to code locations
where changes are required, as is done with any ill-typed Java programs. We now review the main evolution
cases.

\myparagraph{Declaring new entities} A new entity can be declared in
the taxonomy at any time. This evolution does not require any code
changes beyond the implementation of the generated abstract
class for the entity and deployment of its instances. 

\myparagraph{Extending an entity} An entity declaration can be
extended with additional functionalities and attributes. As in a class
hierarchy, these extensions do not require any changes in the existing
code besides the implementation of these new functionalities.

\myparagraph{Removing an entity or one of its functionalities} An
entity declaration or an entity's functionality can be removed by the
area expert. Such a change requires the modification of the
architecture and the regeneration of the programming
framework. The new programming framework triggers Java
compile-time errors in the implementation code. These errors concern the entity
developers and potentially the application developers.

\subsection{Evolution of the Architecture}

Similarly, an architect may want to change the context and controller
component descriptions of the architecture. Again, this requires a
regeneration of the programming framework that can lead to
compile-time errors to be resolved by the application developer.

\myparagraph{Adding a new context or controller} New context and
controller declarations can be added. This does not require any code
changes beyond the implementation of the newly generated abstract
classes.

\myparagraph{Adding a new input source} A context or controller
declaration can be added a new input source or context, solely
requiring the implementation of the newly generated abstract methods.

\myparagraph{Removing an input source} A context or controller
declaration can be removed an input source or context. In this case,
code that deals with this input source becomes dead. The Java compiler
detects this situation and reports it as an error, requiring the
developer to remove the dead code.\footnote{This feature depends on
  the \ct{@Override} annotation introduced in Java 5.}

\myparagraph{Changing a context type} A context type can be changed.
In this case, the implementation of this context as well as
implementations of subscribers of this context have to be fixed
accordingly. Again, the Java compiler displays meaningful errors
to guide the developer.

\myparagraph{Adding or removing actions} A controller can be added or
removed device actions. Removal of a device action in a controller
leads to compile time errors where the action is used. An action
added to a controller does not affect the controller implementation in
any way, but new methods are available to be used.

\subsection{Changing the Deployed System}

Our system supports basic runtime changes.

\myparagraph{Plugging and unplugging entity instances} New entity
instances can be added, or removed without requiring any
changes in the application code. New instances are registered into the
framework and become immediately available via entity discovery.
Unplugged entities are detected and automatically made unavailable to
entity discovery.

\myparagraph{Implementing and plugging new entity implementations} An
entity developer can implement a new kind of entity without requiring
any changes. A system administrator can deploy new instances of this
implementation without restarting the system.

\medskip

This section presented how our approach supports changes in the
taxonomy and architecture, late in the development process. \final{To
  further enhance this process, we believe we could provide the
  architect and area expert with refactoring tools to ease changes in
  developer's code.} This could be done in different ways, the
simplest being the generation of migration documentation. More
challenging solutions could propagate refactorings from the
architecture and taxonomy to the implementation.

\section{Evaluation of our methodology}
\label{sec:evaluating}

In this section, we conduct an evaluation of our methodology. To do so,
we explore three aspects: (1)
\textit{expressiveness}, evaluating the scope of this methodology, (2)
\textit{usability}, estimating the intuitiveness of the tools, and (3)
\textit{productivity}, measuring development time, code quality, and
reusability.

\subsection{Expressiveness}

We study the expressiveness of the \diaspec{} specification language
by evaluating the scope of the underlying architectural pattern. This
expressiveness is evaluated by developing
a wide range of applications in multiple areas.
We now describe some of them.

\paragraph*{In-lab deployment}

To show that our framework is operational, we have deployed several
applications in a dedicated room of our lab. The first application is
an anti-intrusion system. It is responsible for securing a room with
password-protected lockers, motion detectors, and alarms. When an
intrusion is detected, the application takes a photo of the intruder
and sends it to a supervisor along with information about the
intrusion~\cite{Cass10a}. The second application is a
  multimedia-content alert system that informs users about their
  preferred TV programs. The third application is a peer-to-peer
  document sharing system that requires user identification through
  various means (\eg{} RFID badges and fingerprint). The fourth
  application is an intranet web-server monitoring prototype. This
  monitoring application logs the profiles of the web server users and
  emails the server administrators in case of intrusion. Finally, the
Newscast application described in this paper has been deployed in our
lab and used in several demos~\cite{Cass10a,Brun09d}.

\paragraph*{A real-size case study} We applied our tool-based
methodology to a real-size case study: the management of a
13,500-square-meters building hosting an engineering school. Six
pervasive computing applications, including the Newscast
  application, were developed for this case study. These
applications cover several pervasive computing areas. The light and
air management applications relate to the building automation area.
The fire management application pertains to the emergency management
area. Finally, the access control and anti-intrusion applications
relate to the security area.

Table~\ref{tab:metrics_casestudy} gives, for each application of this
case study, the number of elements for each type of declarations:
entity classes, context components, and controller components. 

\begin{table}
  \centering
  \scriptsize{}  
  \begin{tabular}{@{}>{\bfseries}l*{5}{c}@{}}
\toprule
\textbf{Application} & \multicolumn{3}{c}{\textbf{Entity}}                 & \textbf{Context} & \textbf{Controller} \\
                     & \textbf{class} & \textbf{source} &  \textbf{action} &                  &\\

\midrule
Newscast         & 7 & 6 & 2 & 6 & 2 \\
Light Management & 5 & 4 & 2 & 3 & 1 \\
Air Management   & 7 & 5 & 4 & 6 & 2 \\
Access Enforcer  & 4 & 5 & 1 & 3 & 1 \\
Anti-Intrusion   & 6 & 4 & 2 & 4 & 2 \\
Fire Management  & 7 & 4 & 3 & 2 & 1 \\
\bottomrule
  \end{tabular}
  \caption{Metrics on our case study}
  \label{tab:metrics_casestudy}
\end{table}

In total, the case study involves 36 classes of entities, 28 data
sources, 14 action definitions, 24 context components, and 9
controllers. The \diaspec{} taxonomy consists of 200 lines of code
(LOC), the architecture 130 LOC, the generated framework 7000 LOC, and
developer-supplied Java code 3000 LOC.

We observe that to cover realistic areas, the number of entity
classes is low: up to 7. This makes the artifacts of our tool-based
methodology manageable for the stakeholders of a development
project; they are an effective vehicle to expose and share design
decisions. This case study showed that most of the
application logic is realized in the context components.
Indeed, the layers of our architecture pattern isolate the
context calculation from its use for controlling the
environment. This layered architecture simplifies the implementation of the
controllers and allows the information processing to evolve
independently from the control. As a consequence, there are few
controllers per area (between one and two), in contrast to the
number of contexts (four, on average).

The engineering school building is simulated using \diasim{}.
In this simulation, over 400 entity
instances and 300 occupants are simulated (\eg{} staff and faculty
members, students, and visitors) with various behavioral patterns.

We are planning on deploying part of the applications in-situ. We have
already started to make unitary tests of equipments and software
entities, thanks to the incremental capability of \diasuite{}.

We notice that our case study was modeled with few
  declarations of entities, contexts, and controllers; and yet this
model scaled up to a rather large simulation scenario, involving
numerous entities, building occupants, and simultaneously running
applications.

\medskip{}

During the development of our approach, we have implemented various
applications covering numerous areas including home/building
automation~\cite{Cass10a,Cass09b}, multimedia adaptation, IP
telephony~\cite{Bert09a,Bert10a}, and health-care~\cite{Drey09a}. The
wide spectrum of areas covered shows the expressiveness of our
methodology in the context of pervasive computing.

\subsection{Usability}

\begin{table}
  \centering
  \scriptsize{}  
  \begin{tabular}{@{}lp{.42\linewidth}ccc@{}}
    \toprule
    \textbf{Phase} & \textbf{Nature of the task}               & \multicolumn{2}{c}{\textbf{\% students}}                 & \textbf{Avg.} \\
                   &                                           & \textbf{full}                            & \textbf{part} & \textbf{time} \\
    \midrule
    Design         & \diaspec{} specification                  & 100\%                                    & 0\%           & 2h            \\
    Implementation & Extension of the generated \textit{fwork.} & 60\%                                     & 40\%          & 5h            \\
    Testing        & \diasim{} simulation                      & 30\%                                     & 0\%           & 1h            \\
    \bottomrule
  \end{tabular}
  \caption{Results of a lab involving 60 undergraduate students.}
  \label{tab:usability}
\end{table}

We have been using \diasuite{} for a course on software architectures for
three years. This course included an 8-hour lab that consisted of twenty groups
of three undergraduate telecommunication students (equivalent to the
master's level) who had never used \diasuite{}. Furthermore, these
students had only followed an introductory course on Java a year
before the software architecture course and had no prior experience in
software design or pervasive computing.

The goal of the lab was to develop a Newscast application from a
diagram similar to that of Figure~\ref{fig:newscast-dataflow}. Using
the diagram, the students did not have to find the decomposition of the
application into \diaspec{} components, simplifying the design stage.
They had to (1) determine appropriate types for the entity sources and
contexts, (2) translate the diagram into \diaspec{}, and (3)
implement the application. We intentionally gave very sparse
information about \diaspec{} and the generated programming framework,
to determine to what extent the language and generated framework were
in themselves able to guide the architecture and development.

The results of this lab are presented in Table~\ref{tab:usability}.
All student groups have managed to design a \diaspec{} architecture in
conformance with the provided diagram. In general, students only
required explanations about the role of each component type (entity,
context, and controller) and the interactions between these types.
During the whole evaluation, \diagen{} produced error messages that
were clear enough to  require no additional information from the
instructor.

At the end of the 8-hour lab, 60\% of the groups managed to develop a
working implementation where all provided unit tests passed
while the remaining 40\% provided a partial implementation where most
of the unit tests passed. 30\% of the students went beyond the
assignment by configuring and testing their application using the
\diasim{} simulator. This first experience demonstrates that
students with modest knowledge in software
engineering are able to efficiently use \diaspec{} in a short
time. 

This first usability evaluation is preliminary. We plan to conduct
empirical usability studies with professional software developers. We believe
that professional software developers with a background in Java and
Eclipse will provide a direct feedback on our approach, without the
interferences observed with our students due to their lack of
acquaintance with the programming tools. Furthermore, we have noticed
that \diasuite{} helps decomposing the development effort into clearly
defined task assignments: each stakeholder needs only a local
knowledge of his task. We would like to conduct further studies on
this aspect to assess its impact in practice.

\subsection{Productivity}

One of the benefits of DSLs is to enhance
productivity~\cite{Kieburtz96}. In the following, we show how our
methodology reduces development time, enhances code quality, and
promotes reusability. We believe these three dimensions give an
insight as to how our methodology improves productivity.

\paragraph*{Development time}

Initial development time is directly proportional to the size of the
written code. Automatic program generation aims to reduce the code to
be written and thus to reduce development time. We measured the
quantity of the generated code in several applications we
developed. Our measures reveal that nearly 80\% of the project code
base is generated, the implementation accounts for 17\%, the rest being
the declarations in \diaspec{}. 

Importantly, these measures are useful only if the generated code is
actually executed. Otherwise the generated code can be arbitrarily
large without impacting development time. We measured the coverage of
the framework code during a number of executions of these
applications, using the CodeCover testing
tool.\footnote{\url{http://www.codecover.org/}} On average, 70\% of
the generated framework is actually executed. We studied the parts of
the framework that are not executed and found that most of them are
either error handling code or features unused by the application
logic.

\paragraph*{Code quality}

A program with a good code quality is a program that evolves
  and is maintained with ease. Code quality is critical as
maintenance of a software system accounts for more than 85\% of the
total cost of an application~\cite{Erli00a}. We measured code quality
of implementations from developers allowing us to assess the
usefulness of our approach in making developers write high quality
implementations. We used
Sonar\footnote{\url{http://www.sonarsource.org/}} to measure the code
quality of three applications: a web-server monitoring application, an
anti-intrusion system, and the Newscast application presented in this
paper. To do so, we used various criteria, as provided by Sonar,
  including code duplication, rules compliance, code coverage, and code
  complexity. As an example, the code complexity is measured using the
  well-known cyclomatic complexity metric as defined by
McCabe~\cite{McCab76a}: this metric measures the number of linearly
independent paths in a source code. The implemented code for the three
projects has an average cyclomatic complexity of 3. McCabe notes that
``code in the 3 to 7 complexity range [...] is quite well
structured''. A complexity greater than 10 indicates very poor
code quality, which hampers maintenance and thus productivity.
The other criteria, as offered by Sonar, present 
  results indicating an overall good quality of written source code.
  These results, associated with the small percentage of code written
manually, reveals that our generated programming framework guides the
developers through a well-structured and easy-to-maintain code.

\paragraph*{Reusability}

\begin{figure}
  \centering
  \includegraphics[width=.8\linewidth]{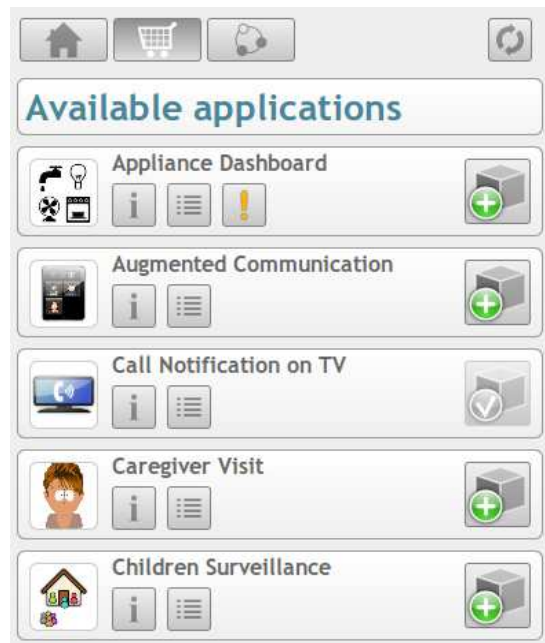}
  \caption{\label{fig:diastore} DiaStore, a web application to share
    and deploy new \diaspec{} applications.}
\end{figure}

Our approach promotes reuse of the specifications and implementations
across applications. An entity specification and its associated
implementation can be packaged for later reuse. To promote cross
application reuse, we have developed DiaStore, a web application that
allows to easily download and deploy new \diaspec{} applications in
the spirit of the Apple's App
Store\footnote{\url{http://www.apple.com/iphone/features/app-store.html}}
(see Figure~\ref{fig:diastore}). Using the declarations from the
taxonomy, DiaStore indicates the entities currently deployed at home
and checks whether the entity requirements of a new application are
fulfilled. This approach encourages the sharing of entities between
applications.

We have also noticed that the \diaspec{} architectural pattern
encourages context reuse. Specifically, intra-application reuse arises
for applications sharing refined sensed data via context components.

This first evaluation of our tool-based methodology is promising. The scope of
the applications embraces most of the pervasive computing area. The
intuitiveness of the generated framework has been validated with students who
managed to develop an entire Newscast application, without any API
documentation, solely using the IDE completion. Finally, the productivity gain
has been evaluated through several criteria measuring development time, code
quality, and reusability.

\section{Related work}
\label{sec:related-works}

In this section, we present a comparison study of the existing
approaches for developing pervasive computing applications. This study
addresses each stage of the development cycle, namely, design,
implementation, testing, deployment and evolution/maintenance. We
characterize the existing development approaches and illustrate each
of them with a representative example.  The results of this comparison
are shown in Table~\ref{tab:comparison}. The degree of support for
each development stage is rated from ``+'' to ``++++'': ``+''
corresponds to low-level support, such as guidelines, whereas ``++++''
corresponds to highly customized support or complete automation of the
task. When no support is provided by an approach, no rating is given.

In the remaining of this section, we first present the existing
development methodologies that provide support for the entire
development cycle. Then, we present the development approaches that
target particular stages of the development cycle.

\begin{table}
  \centering
  \footnotesize{}
  \begin{tabular}{l@{~~}*{5}{c@{~~}}}
    \toprule
    ~                                    & \textbf{Design.} & \textbf{Implem.} & \textbf{Test.} & \textbf{Deploy.} & \textbf{Evol./Mainte.} \\
    \midrule
    \textit{DiaSuite}                    & ++++             & +++              & ++             & +++              & +++                    \\ 
    \rowcolor[gray]{.8}   
    \textit{PervML}~\cite{Serr10a}       & ++               & ++++             & +              & ++++             & ++++                   \\
    \textit{ArchFace}~\cite{Ubay10a}     & ++               & ++               &                &                  &                        \\
    \rowcolor[gray]{.8}    
    \textit{Context Tk.}~\cite{Dey01b}   & +                & ++               &                &                  &                        \\
    \textit{Olympus}~\cite{Rang05a}      & +                & ++               &                & +++              &                        \\
    \rowcolor[gray]{.8}    
    \textit{Player/Stage}~\cite{Coll05a} &                  & ++               & ++             & +                &                        \\
    \bottomrule
  \end{tabular}
  \caption{Comparison of the support provided by representative approaches throughout the development cycle. }
  \label{tab:comparison}
\end{table}

\subsection{Model-Driven Engineering}

Model-Driven Engineering (MDE) uses models and model transformations
to specify software architectures and generate
implementations~\cite{Schm06a}. The goal of these MDE approaches is to
raise the abstraction level in program specifications and generate
a working implementation from such a specification. UML~2.0 (Unified
Modeling Language) has been widely accepted as an architecture
modeling notation~\cite{Booc05a} and as a second-generation
ADL~\cite{Medv07a}.
Various development environments, relying on UML and MDE, have been
proposed (\eg{} Enterprise Architect~\cite{EnterpriseArchitect}).
These development environments cover the complete development
life-cycle. However, they do not target the specific features of
pervasive computing, leaving the customization work to the architects
and the developers.

PervML~\cite{Serr10a}, listed in Table~\ref{tab:comparison},
customizes the MDE approach with respect to the domain of pervasive
computing by proposing a conceptual framework for context-aware
applications. This conceptual framework relies on UML diagrams to
model pervasive computing concerns. For example, services are modeled
with class, sequence, and state transition diagrams, while locations
are modeled with package diagrams. Even though the conceptual
framework proposed by PervML is domain specific, it relies on generic
notations and generic tools, incuring an overhead for designers.

In contrast, \diaspec{} designers only manipulate domain-specific
concepts and notations (\eg{} entities, context and controller
components), facilitating the design phase. PervML, along with most
MDE-based approaches, require designers to directly manipulate OCL and
UML diagrams. As reported in the literature, this manipulation becomes
``enormous, ambiguous, and
unwieldy''~\cite{Picek08a,Fowl03b,Thom04b}. In practice, these
approaches demand an in-depth expertise in MDE technologies. For the
design phase, PervML is rated ``++''.

From UML diagrams, PervML provides a dedicated suite of tools to
generate a complete implementation. This approach is thus rated
``++++'' for the implementation stage. Using UML diagrams allows to
leverage developers' knowledge and existing tools, such as the Eclipse
Graphical Modelling Framework~(GMF) and the OSGi deployment model; it
is thus rated ``++++'' for the deployment. 

PervML offers rudimentary testing support, based on device simulation;
this phase is rated ``+''.  By leveraging MDE development
environments, the evolution of an existing PervML application only
requires to modify its model and to re-generate the
implementation. This evolution capability results in rating the
corresponding PervML stage ``++++''.

\subsection{Architecture Description Languages}

Architecture Description Languages (ADLs) are used to make explicit
the design of an application. Most ADLs are dedicated to analyzing
architectures; they provide little or no implementation
support. Archface~\cite{Ubay10a} is the most recent instance of this
line of work (Table~\ref{tab:comparison}). It is both a
general-purpose ADL and a programming-level interface. It proposes an
interface between design and code.  However, the design support
provided by Archface is generic. Furthermore, Archface requires the
software architect to have some knowledge about the implementation
layer to be able to express the interface part of a design.

In contrast, our approach is domain specific and thus allows domain
experts to design their architecture without implementation
knowledge. The design is then used to generate dedicated programming
support for the developer. For example, \diagen{} generates dedicated
programming support to discover entities based on the taxonomy
definition. In Archface, a design is directly mapped into
programming-level interfaces, ensuring the conformance between the
design and the implementation. However, unlike our approach, Archface
does not provide dedicated programming support. Taking into account
these limitations, Archface is rated ``++'' for both the design
and implementation stages. The other development stages are not
covered by this approach.

\subsection{Context management middlewares}

Numerous middlewares have been proposed to support the implementation
of pervasive computing applications. Schmidt \etal{}~\cite{Schm99b},
Chen and Kotz~\cite{Chen02b}, and Dey \etal{}~\cite{Dey01b} have
proposed middleware layers to specifically acquire and process context
information from sensors. The Context Toolkit proposed by
Dey~\cite{Dey01b} illustrates this approach in
Table~\ref{tab:comparison}. Henricksen \etal{} take this approach one
step further by introducing a language to model the computation of
context information~\cite{Henr04a,McFad05a}. However, none of these
middlewares provide tool support for the design phase, they only
provide design guidelines; Context Toolkit is thus rated ``+''.

Although, context management middlewares provide programming support
for acquiring and processing context information from sensors, they do
not address the other activities pertaining to a pervasive computing
application (\eg{} device actuation). Because of this limitation, the
programming support of this approach is rated ``++''.  The other
development stages are not covered by these middlewares.

\subsection{Programming Frameworks}

The programming framework approach has been applied to the domain of
pervasive computing to facilitate the development of applications by
raising the level of abstraction. A representative example is
Olympus~\cite{Rang05a}, included in
Table~\ref{tab:comparison}. Olympus offers limited support for the
design stage: it mainly consists of guidelines related to the concept
of Active Space. This stage is rated ``+''.

An Active Space represents a physical space enriched with sensors and
actuators. Virtual entities of an Active Space can be described
programmatically using high-level programming interfaces, allowing the
developer to focus on the application logic. However, the programming
support is not dedicated to a specific description of an Active
Space. Thus, the application logic is implemented using generic
datatypes, making the implementation error-prone. In comparison,
DiaSuite provides datatypes to the developer that are dedicated to the
application to be implemented. Olympus implementation support is rated
``++''. 

The high-level nature of the description of an Active Space allows to
reuse the same program across pervasive computing environments, easing
the deployment stage; it is thus rated ``++++''.

\subsection{Simulators}
\label{sec:relatedwork-testing}

Few simulators are dedicated to the testing of pervasive computing
applications. Stage and Gazebo are simulators dedicated to the Player
programming framework and have been used to simulate a sensor-enriched
kitchen~\cite{Kran06a}. Player is a programming framework and a
middleware created in the robotics domain and widely recognized as a
standard for robot programming~\cite{Coll05a}.  

Other pervasive computing simulators include Ubiwise~\cite{Ubiwise}
and Tatus~\cite{ONeil05a} that are built upon 3D first-person
game-rendering engines; they allow the user to have a focused
experience of a simulated environment. However, these simulators are
difficult to extend: the game-rendering engine has to be modified to
add new sensors and actuators, or to simulate arbitrary context
data. The PiCSE simulator addresses the problem of extensibility by
providing generic libraries to create sensors and
actuators~\cite{Reynolds06}.

In Table~\ref{tab:comparison}, the Stage simulator combined with the
Player programming framework illustrate this compound approach.
Player allows to specify interfaces that define how to interact with
robotic sensors, actuators and algorithms. However, this design
support is very limited as it does not cover the design of other
application components. For instance, it does not allow to design the
controllers that coordinate robotic devices.  Thus, Stage is rated
``+'' for the design phase. 

The Player programming support enables to develop a wide range of
robotic applications. However, this programming support only targets
the robotic area, resulting in a ``++'' rating. Player applications
can be simulated in a 2D graphical environment using Stage, or in 3D
using Gazebo. However, both simulators only target the simulation of
mobile robots; their testing support is rated ``++''. These simulators
provide guidelines to deploy applications but do not address
evolution; they are rated ``+'' for the deployment stage.  Moreover,
they have to be manually specialized for every new application area.
In contrast, \diasim{} relies on the \diaspec{} descriptions to
automatically customize the simulation tools (\ie{} the editor and the
renderer).

\subsection{Summary}

The comparison in Table~\ref{tab:comparison} first shows that
\diasuite{} provides a comprehensive support for the design and
programming stages, compared to the existing approaches. It also shows
the support for the testing, deployment, and evolution stages can be
improved. To do so, we plan on leveraging existing technologies that
focus on these stages. For example, to improve the deployment and
runtime evolution support, we are working on leveraging OSGi, as is
done by PervML.

\section{Conclusion}
\label{sec:conclusion}

In this paper, we  presented \diasuite{}, a tool-based methodology
for developing real-size pervasive computing applications. Our
methodology provides support throughout the development life-cycle of a
pervasive computing application: design, implementation, simulation,
and execution. First, the taxonomy of the target area and the
architecture descriptions are written in the \diaspec{} language.
Then, the \diagen{} compiler processes these descriptions and
generates a dedicated programming framework. This framework raises the
abstraction level by providing the programmer with high-level
operations for entity discovery and component interactions. A
pervasive computing simulator, \diasim{}, is used to simulate the
environment. Finally, the \diasuite{} back-end enables to deploy an
application using a specific distributed systems technology.

Our methodology has been successfully applied to the
development of realistic pervasive computing applications in a wide
spectrum of areas. The evaluation of our methodology has demonstrated
its benefits for every stage of the development life-cycle.

\subsection*{Assessments}

We now assess our tool-based methodology with respect to our initial
objectives.

\myparagraph{Heterogeneity} Our taxonomical approach has been
successful at taming the heterogeneity of devices and software
components. This is demonstrated by the spectrum of entities modeled
to cover the areas of our case study and the ease at implementing
entities from their declarations. This approach also showed to be
effective for reusing entity declarations across areas.

\myparagraph{Architecture} Decomposing an application into contexts and
controllers has been a useful process to identify the key processing
units of an application. This decomposition was found to greatly
simplify the implementation phase.

\myparagraph{Development cycle} The \diaspec{} design language has
proved to be a great asset to introduce developers to the pervasive
computing domain. We were able to validate this benefit during a
course on software architectures in which the students were asked to
develop an application with \diasuite{}. \diaspec{} was instrumental
in providing them with a conceptual framework to develop their
application, with artifacts to get feedback and programming support
from \diaspec{}-processing tools.

\myparagraph{Simulation} \diasim{} has been an essential instrument to
validate our tool-based methodology. Every part of \diasim{} is
customized with respect to a \diaspec{} description. Another benefit
of our generative approach is that it allows hybrid environments,
combining simulated and real entities. \diasim{} makes it possible to
tackle real-size pervasive computing applications, without testing
them on toy platforms or undertaking extensive deployment of
equipments.

\subsection*{Ongoing and future work}

This work is being expanded in various directions.

\myparagraph{Towards a visual design language} Visual design languages
such as UML improve the readability and usability of design
descriptions, promoting design-driven development. We have started
working on a graphical notation for \diaspec{}. This notation relies
on the layered data-flow view of a \diaspec{} architecture, as shown
in Figure~\ref{fig:newscast-dataflow}.

\myparagraph{Describing component interactions} Mapping a software
architecture to an implementation is a well-known
challenge~\cite[Chap.~9]{Tayl09a}. A key element of this mapping is
the architecture's description of the data and control-flow
interactions between components. We have introduced a notion of
\textit{interaction contract} that expresses the set of allowed
interactions between components, describing both data and control-flow
constraints~\cite{Cass11a}. This declaration is part of the
architecture description, allows generation of extensive programming
support, and enables various verifications.

\myparagraph{Adding non-functional layers} One direction consists of
widening the scope of \diasuite{} by introducing non-functional
concerns (\eg{} fault-tolerance, safety, and security) in our
tool-based methodology. Work is in progress to add and compose
non-functional layers on top of the \diaspec{} language and have
automatically generated programming support~\cite{Jakob09a}. For
example, a safety expert may want to specify at design time how errors
are handled, guiding and facilitating the implementation of error
handling code~\cite{mercadal2010}. We have also added Quality of
Service (QoS) declarations to \diaspec{} to ensure QoS requirements
traceability through every stage of the development
life-cycle~\cite{Gatt11a}. Composition of non-functional layers is
difficult and raises issues similar to aspect
composition~\cite{Sanen08a,Doue02b}.

\myparagraph{Enhancing simulation} Simulating natural phenomena like
heat propagation can be quite complex as they involve mathematical
equations. We are actively working on easing simulation of these
phenomena by leveraging Acumen, a DSL for describing differential
equations~\cite{Zhu09a}. The differential equations defined with
Acumen describe physical phenomena. We would also like to add
contracts to \diaspec{} in the form of pre- and post-conditions to
entities, controllers, and contexts. These contracts could drive the
rendering of a simulation by drawing the tester's attention when they
are not met. This is particularly useful for large-scale simulations
in which numerous events occur at the same time, complicating the
monitoring of a simulation. With these contracts, the tester is able
to specify the important events. For example, an alert should be
raised when a door is locked while the building is on fire. We could
also use these contracts to inform the system administrator about
potential problems.

\myparagraph{Verification} Another promising direction is to take
advantage of architectural invariants for guiding program analysis
tools. Our generative approach could automatically add architectural
invariants as axioms to the model, facilitating verification. For
example, we are investigating the verification of safety properties by
injecting the architectural invariants from the \diaspec{}
specification in the model checker JPF~\cite{Visser00}.

\myparagraph{Empirical evaluation} The evaluation presented in
Section~\ref{sec:evaluating} is preliminary; we plan to conduct an
empirical evaluation based on a well-defined experimental methodology.
In particular, we would like to evaluate the usability and
productivity gained by comparing our approach with existing tool-based
development methodologies for pervasive computing applications.

\bibliographystyle{IEEEtran}

\end{document}